\documentclass[11pt,a4paper]{article}
 \usepackage{jheppub}
\usepackage{graphicx,epsfig,psfrag,amssymb,hyperref}
\usepackage{multirow}
\usepackage{color,xcolor,graphicx,epsfig,psfrag,amsmath,empheq}
\usepackage{bm}
\usepackage{mathrsfs,amsfonts,, color}
\usepackage[caption=false]{subfig}
\usepackage{hepunits}
\usepackage{color}
\usepackage[utf8]{inputenc}
\definecolor{darkgreen}{rgb}{0,0.5,0}

\makeatletter
\def\@seccntformat#1{\csname the#1\endcsname.\quad}

\def\lsim{\mathrel{\raise.3ex\hbox{$<$\kern-.75em\lower1ex\hbox{$\sim$}}}}
\def\gsim{\mathrel{\raise.3ex\hbox{$>$\kern-.75em\lower1ex\hbox{$\sim$}}}}

\newcommand{\be}{\begin{eqnarray}}
\newcommand{\ee}{\end{eqnarray}}
\newcommand{\Br}{\textrm{Br}}

\newcommand{\Tr}{\textrm{Tr}}

\newcommand{\UD}{{ U(1)}_D}
\newcommand{\URH}{{ U(1)}_{RH}}

\newcommand{\ga}{\gamma}
\newcommand{\apr}{A^\prime}

\newcommand\FNAL{Fermi National Accelerator Laboratory, \\ Batavia, IL USA}
\newcommand\Princeton{Princeton University, \\ Princeton, NJ USA}
\newcommand\UCI{University of California, Irvine, \\ Irvine, CA USA}

\preprint{\\ FERMILAB-PUB-16-385-PPD, UCI-HEP-TR-2016-15, MITP/16-098, PUPT 2507}

\title{Light Weakly Coupled Axial Forces: Models, Constraints, and Projections}

\author[a]{Yonatan Kahn,}
\author[b]{Gordan Krnjaic,}
\author[a]{Siddharth Mishra-Sharma,} 
\author[c]{and Tim M.P. Tait}

\affiliation[a]{\Princeton}
\affiliation[b]{\FNAL}
\affiliation[c]{\UCI}

\emailAdd{ykahn@princeton.edu}
\emailAdd{krnjaicg@fnal.gov}
\emailAdd{smsharma@princeton.edu}
\emailAdd{ttait@uci.edu}

\abstract{
We investigate the landscape of constraints on MeV-GeV scale, hidden $U(1)$ forces with nonzero axial-vector couplings to Standard Model fermions. While the purely vector-coupled dark photon, which may arise from kinetic mixing, is a well-motivated scenario, several MeV-scale anomalies motivate a theory with axial couplings which can be UV-completed consistent with Standard Model gauge invariance. Moreover, existing constraints on dark photons depend on products of various combinations of axial and vector couplings, making it difficult to isolate the effects of axial couplings for particular flavors of SM fermions. We present a representative renormalizable, UV-complete model of a dark photon with adjustable axial and vector couplings, discuss its general features, and show how some UV constraints may be relaxed in a model with nonrenormalizable Yukawa couplings at the expense of fine-tuning. We survey the existing parameter space and the projected reach of planned experiments, briefly commenting on the relevance of the allowed parameter space to low-energy anomalies in $\pi^0$ and $^8{\rm Be}^*$ decay.}

\begin{document}
\maketitle

\section{Introduction}

New sub-GeV abelian gauge bosons are simple, well-motivated extensions of the Standard Model (SM). If SM particles are singlets under the corresponding $U(1)_D$ ($D$ for ``dark'') group, the leading SM interaction with the new gauge boson arises through kinetic mixing with the hypercharge field strength tensor, such that the associated dark gauge boson $A'$ couples predominantly to the electromagnetic current after electroweak symmetry breaking (EWSB) \cite{Okun:1982xi,Holdom:1985ag}. 
Alternatively, if $U(1)_D$ gauges a subset of SM quantum numbers, the new gauge boson couples directly to a current of SM  fields, which can radiatively induce a nonzero kinetic mixing as well; popular examples include the anomaly-free combinations $B-L$ \cite{Hewett:1988xc, Khalil:2006yi,Khalil:2007dr}, $L_i - L_j$ \cite{Altmannshofer:2014cfa,Altmannshofer:2014pba}, $B-3L_i$ \cite{Ma:1997nq}, and $B-L + x Y$ \cite{Lee:2016ief}, where $x \in \mathbf R$. These abelian extensions are ubiquitous in the model-building literature and regularly invoked to explain anomalies in dark matter detection \cite{Boehm:2003hm,Boehm:2003bt,Boehm:2004uq} and resolve discrepancies in precision physics measurements \cite{Feng:2016jff,Kahn:2007ru}, to name only a few applications.

However, these extensions typically induce sizable $A^\prime$ interactions only with vector currents of SM fermions, which limits their applicability in phenomenological settings that also require axial couplings (\emph{e.g.} parity violating observables). In this paper, we explore the expanded parameter space of light, weakly-coupled abelian gauge bosons with both axial and vector couplings. We are motivated in part by the observation in \cite{Kahn:2007ru} that an axially-coupled vector $A'$ could contribute at tree-level to the rare decay $\pi^0 \to e^+ e^-$, which is loop- and helicity-suppressed in the SM, to resolve a $2-3\sigma$ discrepancy between theory \cite{Dorokhov:2007bd} and experiment \cite{Abouzaid:2006kk}. Recent work \cite{Masjuan:2015lca,Masjuan:2015cjl} parameterizing the amplitudes for this decay and the similar $\eta, \eta' \to e^+ e^-, \, \mu^+ \mu^-$ processes has made it possible to test hypothetical new physics contributions to these rare pseudoscalar decays against SM predictions. Such contributions necessarily involve axial couplings to both quarks 
 and leptons. Furthermore, the $A'$ mass and couplings which resolve the $\pi^0$ discrepancy are similar in magnitude to those which could explain a recent anomaly in decays of an excited state of $^8{\rm Be}$ \cite{Krasznahorkay:2015iga,Feng:2016jff,Feng:2016ysn}. An axially-coupled $A'$ could plausibly contribute to the landscape of models relevant for this observation. Other recent work investigating  MeV-scale chiral forces includes \cite{Harigaya:2016rwr}, which builds a light chiral dark sector and  \cite{Correia:2016xcs}, which considers muon specific interactions; this work takes a generic approach to such forces, which
 can be adapted to various circumstances.

The most general Lagrangian for a massive gauge boson $A'$ with both vector and axial couplings is
\be
\label{eq:AprimeLag}
\mathcal{L}_{A'} = -\frac{1}{4}F'_{\mu \nu}F'^{\mu \nu} - \frac{m_{A'}^2}{2}  A'_\mu A'^{\mu} + A'_\mu  \sum_{f } \bar f \left(  c_V^f   \gamma^\mu  + c_A^f   \gamma^\mu \gamma^5  \right)  f,
\ee
where $F'_{\mu \nu} = \partial_\mu A^\prime_\nu-\partial_\nu A^\prime_\mu$ is the 
field strength tensor, $f$ is a (four-component) SM fermion and   $c_{V,A}^{f}$ are its vector and axial-vector couplings. 
It is worth mentioning that even a conventional kinetically-mixed $A'$ will also have small axial couplings suppressed by $m^2_{A'}/m^2_Z$ compared to its vector coupling to the SM electromagnetic current. These are inherited from mixing with the $Z$, since in the UV the $A'$ field strength must be mixed with unbroken hypercharge. For the remainder of this paper we will focus on the case of comparable vector and axial couplings, in contrast with  the typically suppressed axial couplings characteristic of kinetic mixing.\footnote{On the other hand, mass mixing (as opposed to kinetic mixing) with the $Z$ can generate comparable vector and axial couplings \cite{Davoudiasl:2012ag,Davoudiasl:2012qa,Davoudiasl:2014kua}; this will play an important role in our models.}

If we make the further assumption that $\UD$ plays no role in flavor breaking, gauge invariance of the SM Yukawa sector introduces nontrivial relationships between IR and UV physics. Maintaining
unsuppressed axial couplings at low energies generically requires:
\begin{itemize}
\item {\bf Extended Higgs Sector:} Non-vanishing axial $\apr$ couplings and gauge invariant SM
Yukawa couplings jointly require the SM Higgs to carry nonzero $\UD$ charge, so after EWSB,
the Higgs vacuum expectation value (VEV) introduces $\apr-Z$ mass mixing. For small mixings, the rotation angle which diagonalizes the mass terms introduces an additional correction to the $\apr$ axial coupling to SM fermions inherited from the $Z$ neutral-current interaction. As we will derive in section~\ref{sec:1HDM}, if there is only one Higgs doublet in the theory, this additional correction cancels all axial interactions to cubic order in the $\UD$ gauge coupling:
\be \label{eq:axial-afterEWSB-1HDM}
   c_A^f \to  c_A^f +\Delta { c_{A}^f} = 0   + \mathcal{O}(g_D^3)  + \mathcal{O}(\hat{m}^2_{\apr}/\hat{m}^2_Z)   ~~~~ ({\rm after~EWSB}).
\ee 
Thus unsuppressed axial couplings require at least an additional SM Higgs doublet.
\item {\bf New Fermions}: Gauging the SM under a new axially coupled $\UD$ typically 
introduces anomalies through $U(1)_D^3$, $U(1)_D^2 U(1)_Y$,  and $U(1)_Y^2 U(1)_D$ triangle diagrams.  Canceling these diagrams requires new fermions with SM charges (``anomalons") and chiral interactions, implying that their masses arise from Yukawa interactions with a dark Higgs.  As we will show in section~\ref{sec:2HDM}, the null results of new colored
fermion searches at the LHC, which requires their masses to satisfy $\gtrsim 1 \ \TeV$ \cite{Khachatryan:2015oba}, combined with perturbative unitarity of anomalon Yukawa couplings to dark Higgses, implies a lower limit on the $A'$ mass:
\be
\label{eq:anomalonconstraintrough}
m_{A'} \gtrsim 80 \ \MeV \times \left(\frac{g_D}{10^{-3}}\right) \times \left(\frac{4\pi}{y_\psi}\right),
\ee
where $g_D$ is the dark gauge coupling and $y_\psi$ is the anomalon Yukawa.
\end{itemize}

The $A'$ may also be a mediator to the dark sector and  couple to a dark matter candidate $\chi$. If $m_\chi < m_{A'}/2$ and the coupling $g_D$ to the dark sector is relatively strong, the phenomenology of the $A'$ changes significantly, and different bounds constrain the available parameter space (for example, heavy pseudoscalar meson decays to $\gamma + {\rm invisible}$ \cite{Fayet:2007ua}). In this paper, we will assume that a light $\chi$ is not in the spectrum, but there may be interesting regions of parameter space where the $A'$ decays invisibly, or where there are sizable branching fractions to both the visible and dark sectors.

This paper is organized as follows. Section \ref{sec:IRConstraints} outlines the types of experimental constraints most relevant for low-energy axial couplings, in the mass range $2m_e < m_{A'} < 2m_\mu$. Section \ref{sec:GaugeInvtYukawas} describes models where axial couplings are determined by gauge invariance of the SM Yukawa sector. In section \ref{sec:1HDM}, we attempt to build a simple model of axial couplings using only the single Higgs doublet of the Standard Model, and demonstrate the cancellation in Eq.~(\ref{eq:axial-afterEWSB-1HDM}) which results in suppressed axial couplings at low energies. However, this model is useful for illustrating some features and relations between couplings in the Lagrangian, Eq.~(\ref{eq:AprimeLag}), which generically arise in a model where $\UD$ does not participate in flavor breaking. In section \ref{sec:2HDM}, we generalize to a two-Higgs doublet model and discuss the particle content required for anomaly cancellation, as well as the conditions for obtaining unsuppressed axial couplings. In section \ref{sec:Survey}, we survey the parameter space of this model, showing how a tuning of parameters in the IR to avoid certain low-energy constraints typically requires other couplings to be present, making additional constraints relevant. Section \ref{sec:NonRen} relaxes the assumption on flavor breaking, and considers the possibilities of either mass mixing or generation of SM Yukawa terms from $\UD$ breaking. Finally, in section \ref{sec:Conclusion} we offer
some concluding remarks. The Appendices contain a taxonomy of the vector and axial coupling dependence of the low-energy constraints we consider, as well as details on the calculation of the $A'$ contribution to pseudoscalar decays to lepton pairs.


                                                               \section{IR Constraints on Axial Couplings}
                                                               \label{sec:IRConstraints}


The phenomenology of an $A'$ with both axial and vector couplings differs from the more familiar case of kinetic mixing in several important respects. Here we survey the most relevant categories of low-energy constraints, leaving a detailed taxonomy of the dependence on the various couplings to appendix \ref{app:Taxonomy}. For the remainder of this paper, we focus on the mass range $2m_e < m_{A'} < 2m_\mu$, but in this section we will also briefly describe the constraints relevant outside this mass range.

\subsection{$g-2$ Constraints}

New vector bosons can contribute to $(g-2)_e$ and $(g-2)_\mu$ in analogy with QED contributions from the SM photon. The vector couplings $c_V^\ell$ contribute positively, while the axial couplings $c_A^\ell$ contribute negatively. The relative size of the axial contribution compared to the vector contribution is proportional to $m_\ell^2/m_{A'}^2$. The most recent measurements show $(g-2)_e$ is consistent with the Standard Model to within $\sim 1 \sigma$, with the small deviation being negative. This results in a relatively weak constraint on $c_A^e$, since the axial contribution is suppressed by $m_e^2/m_{A'}^2$ compared to the vector contribution, and the data accommodate a small negative contribution; $c_V^e = 0$ but $c_A^e \neq 0$ is allowed for sufficiently large $m_{A'}$. In sharp contrast, $(g-2)_\mu$ has a persistent $3\sigma$ positive deviation between theory and experiment, so a large $c_A^\mu$ is severely constrained unless accompanied by an even larger $c_V^\mu$, or some other (positive) new physics contribution to $(g-2)_\mu$.

\subsection{Parity Violation Constraints}

An $A'$ with both vector and axial couplings $c_A$ and $c_V$ can contribute to parity-violating observables, which are proportional to $c_A c_V$. These constraints are not generally relevant for a kinetically-mixed $A'$ with suppressed axial couplings. For an MeV-scale $A'$, observables measured at extremely low momentum transfer $Q^2 \ll (1 \ \GeV)^2$ are the most stringent since they scale as $1/m_{A'}^2$. Constraints measured at larger $Q^2$ may be relevant, but are approximately independent of $m_{A'}$ for $m_{A'}^2 \ll Q^2$.
\begin{itemize}
\item {\bf Atomic parity violation:} Measurements of the weak charge of cesium constrain the product $c_V^q c_A^e$. The agreement between experiment and theory is at the level of $10^{-5}$ \cite{Porsev:2009pr}, and the measurement is taken at $Q^2 \sim (30 \ \MeV)^2$, making this observable the most sensitive probe of electron axial couplings for a light $A'$. However, this constraint disappears if the $A'$ has vanishing vector couplings to first-generation quarks, $c_V^{u,d} = 0$.
\item {\bf Parity-violating M{\o}ller scattering:} The most precise measurement of parity-violating M{\o}ller (electron-electron) scattering comes from E158 at SLAC \cite{Anthony:2005pm}, at $Q^2 \sim (100 \ \MeV)^2$. This constrains the product $c_V^e c_A^e$.
\item {\bf Neutrino-electron scattering:} A kinetically-mixed dark photon will have no tree-level couplings to neutrinos, and will acquire suppressed couplings through mixing with the $Z$. As we will see in the following sections, gauge invariance in the lepton sector usually ties nonzero lepton axial couplings to neutrino couplings. Neutrino-electron scattering constrains a complicated combination of $c_V^e c^\nu$ and $c_A^e c^\nu$ due to interference of the $A'$ amplitudes with the $Z$ amplitudes (see \cite{Bilmis:2015lja,Jeong:2015bbi} for a complete analysis). The most stringent constraints are from Borexino \cite{Bellini:2011rx} for $\nu_e-e$ and TEXONO \cite{Deniz:2009mu} for $\bar{\nu}_e-e$, both at $Q^2 \sim (1 \ \MeV)^2$, and CHARM-II \cite{Vilain:1994qy} for $\bar{\nu}_\mu-e$, at $Q^2 \sim (100 \ \MeV)^2$.
\end{itemize}

\subsection{Collider and Beam Dump Constraints}

Just like a kinetically-mixed dark photon, an $A'$ with axial couplings can be produced in electron-positron or proton-proton colliders, as well as electron- or proton-beam fixed-target experiments or beam dumps. The phenomenology of the axial coupling depends primarily on the production mechanism, which may suffer an axial suppression for non-relativistic kinematics. In contrast, the $A^\prime$ partial decay width into a fermion $f$  is proportional to the combination $(c_V^f)^2 + (c_A^f)^2$, which is not uniquely sensitive to the axial coupling in isolation.
\begin{itemize}
\item {\bf Meson decay:} A typical production mechanism for dark photons is pseudoscalar meson decay, $\pi^0, \eta \to \gamma A'$, since the $A'$ can replace the photon in any kinematically-allowed process, and neutral pions are copiously produced in proton-beam experiments. In general, though, this decay is only relevant if the conserved current associated with the $A'$ has a mixed anomaly with electromagnetism and the relevant current of flavor $SU(3)$. This is trivially true for a kinetically-mixed dark photon, since the current is just proportional to the electromagnetic current, but for a general group $\UD$, this decay is highly suppressed if $c_V^{u,d} = 0$, Thus, an $A'$ with purely axial couplings to up and down quarks is not effectively constrained by experiments like NA48/2 \cite{Batley:2015lha}, where $A'$ production arises exclusively from pseudoscalars. Similarly, the region of parameter space constrained by proton beam dumps such as U70 \cite{Blumlein:1990ay} or neutrino experiments such as CHARM \cite{Bergsma:1983rt,Bergsma:1985is}, where the dominant production mechanism is through pseudoscalar decay, will be smaller. KLOE \cite{::2016hdx} also probes the vector meson decay $\phi \to \eta A'$, which may receive additional contributions from axial couplings to quarks because the decay is $s$-wave \cite{Feng:2016ysn,KozaczukTalk}, but estimating this contribution is beyond the scope of this work.

\item {\bf Annihilation or bremsstrahlung:} If the $A'$ is produced through annihilation, $e^+ e^- \to \gamma A'$, or bremsstrahlung, $e \to e + A'$ or $p \to p + A'$, the cross section is generally proportional to $c_V^2 + c_A^2$ for relativistic kinematics. Thus the constraints on a kinetically-mixed dark photon are qualitatively similar to those for a dark photon, using $\epsilon e \to \sqrt{c_V^2 + c_A^2}$ to translate between kinetic mixing $\epsilon$ and axial and vector couplings.
\item {\bf Other production mechanisms:} A recent proposal \cite{Ilten:2015hya} suggests exploiting the large flux of $D^*$ mesons at LHCb to search for $D^* \to D + A'$, or more generally performing an inclusive search using data-driven techniques to relate any process with a photon to the corresponding dark photon process \cite{Ilten:2016tkc}. It is difficult to analyze the effect of axial couplings for such searches in general terms, since the hadronic matrix elements may have complicated dependence on axial currents which are not present in the case of pure kinetic mixing. We leave an analysis of such constraints to future work.
\end{itemize}

\subsection{Above $2m_\mu$ and Below $2m_e$}

For $m_{A'} > 2m_\mu$, the possibility of seeing a dimuon resonance in collider experiments or rare meson decays severely constrains the parameter space for axial couplings to quarks, $c_A^q$. For example, vector meson decays $\psi \to \gamma A'$ and $\Upsilon \to \gamma A'$ proceed at tree-level and are proportional to the axial coupling $c_A^{c, b}$ of the heavy quarks \cite{Fayet:2007ua}. If the $A'$ also couples to muons through either an axial or vector coupling, there are stringent bounds on $\Br(\Upsilon \to \gamma \mu^+ \mu^-)$ and $\Br(\psi \to \gamma \mu^+ \mu^-)$ through a resonance. Constraints from BaBar, Belle, and the LHC only depend on the combination $c_V^2 + c_A^2$, and so as mentioned above are qualitatively similar to a kinetically-mixed dark photon with purely vector couplings.

For $m_{A'} < 2m_e$, and assuming no dark sector particles lighter than $2m_{A'}$, the $A'$ can only decay into neutrinos or three photons. For a kinetically-mixed $A'$, both processes have extremely small widths, and the $A'$ can be effectively stable on large timescales. This leads to stringent stellar cooling bounds (see \emph{e.g.}\ \cite{Redondo:2013lna,Vinyoles:2015aba}), where photons in the stellar plasma can convert to $A'$s which carry away energy. However, an $A'$ with axial couplings has drastically different phenomenology, both because it can naturally have unsuppressed couplings to neutrinos, and because the dominant source of $A'$ production in stars is not necessarily through kinetic mixing. A detailed survey of the constraints on an ultralight $A'$ with axial couplings requires a dedicated analysis which is beyond the scope of this work.\footnote{We note, however, that portions of this parameter space may be of great interest to direct-detection experiments searching for keV dark matter, since an eV-scale $A'$ with $c_A^e \sim c_V^e$ can avoid in-medium effects in superconductors which suppress scattering through a kinetically-mixed $A'$ \cite{Hochberg:2015fth,ZurekPrivate}; we discuss this case further in section \ref{sec:Survey}.}

\subsection{IR-Motivated Parameter Space}
\label{sec:IRParameterSpace}

\begin{figure}[t!] 
\hspace*{-0.75cm}
\includegraphics[width=0.502\textwidth]{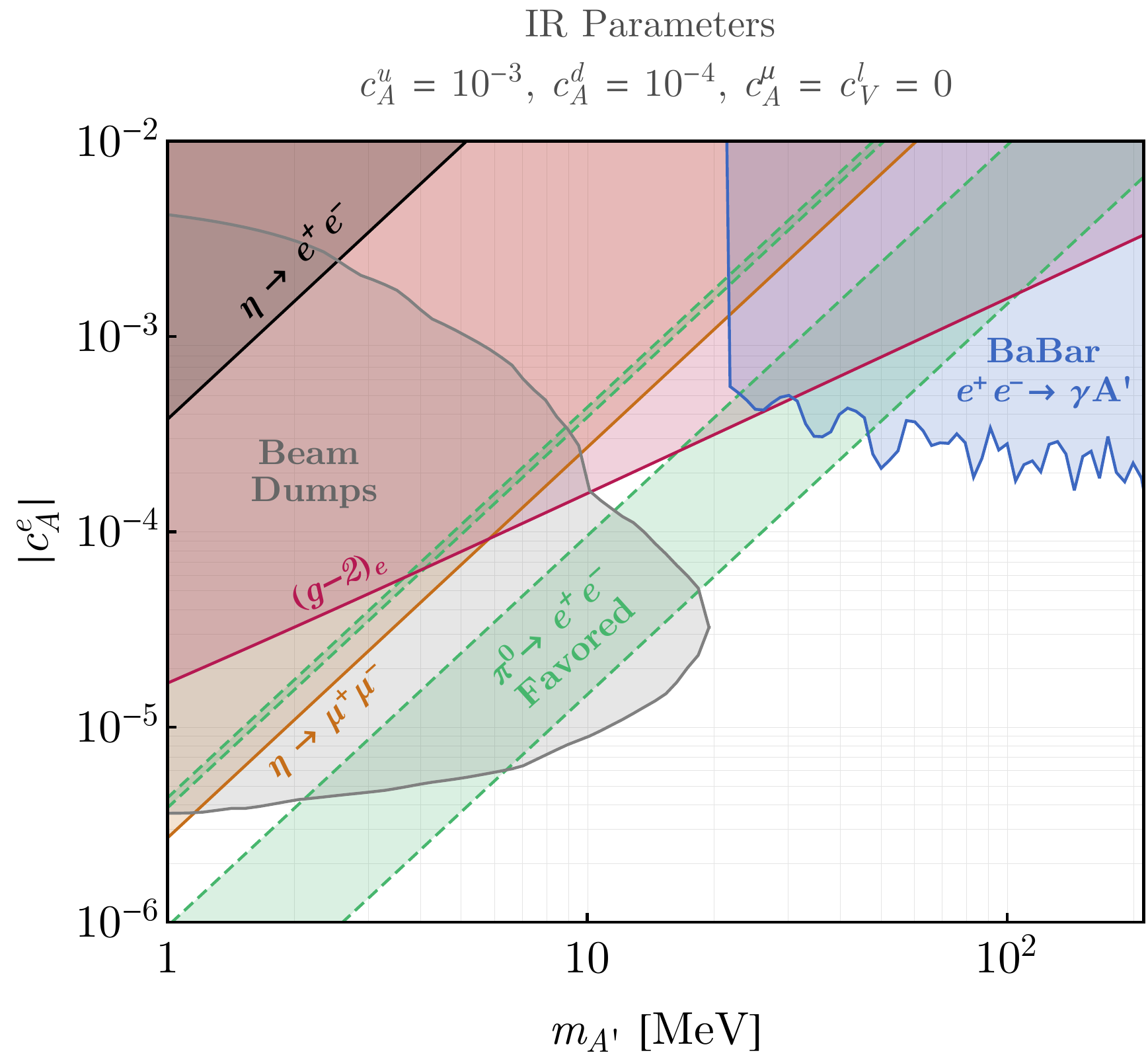} \qquad \hspace*{-0.5cm}
\includegraphics[width=0.504\textwidth]{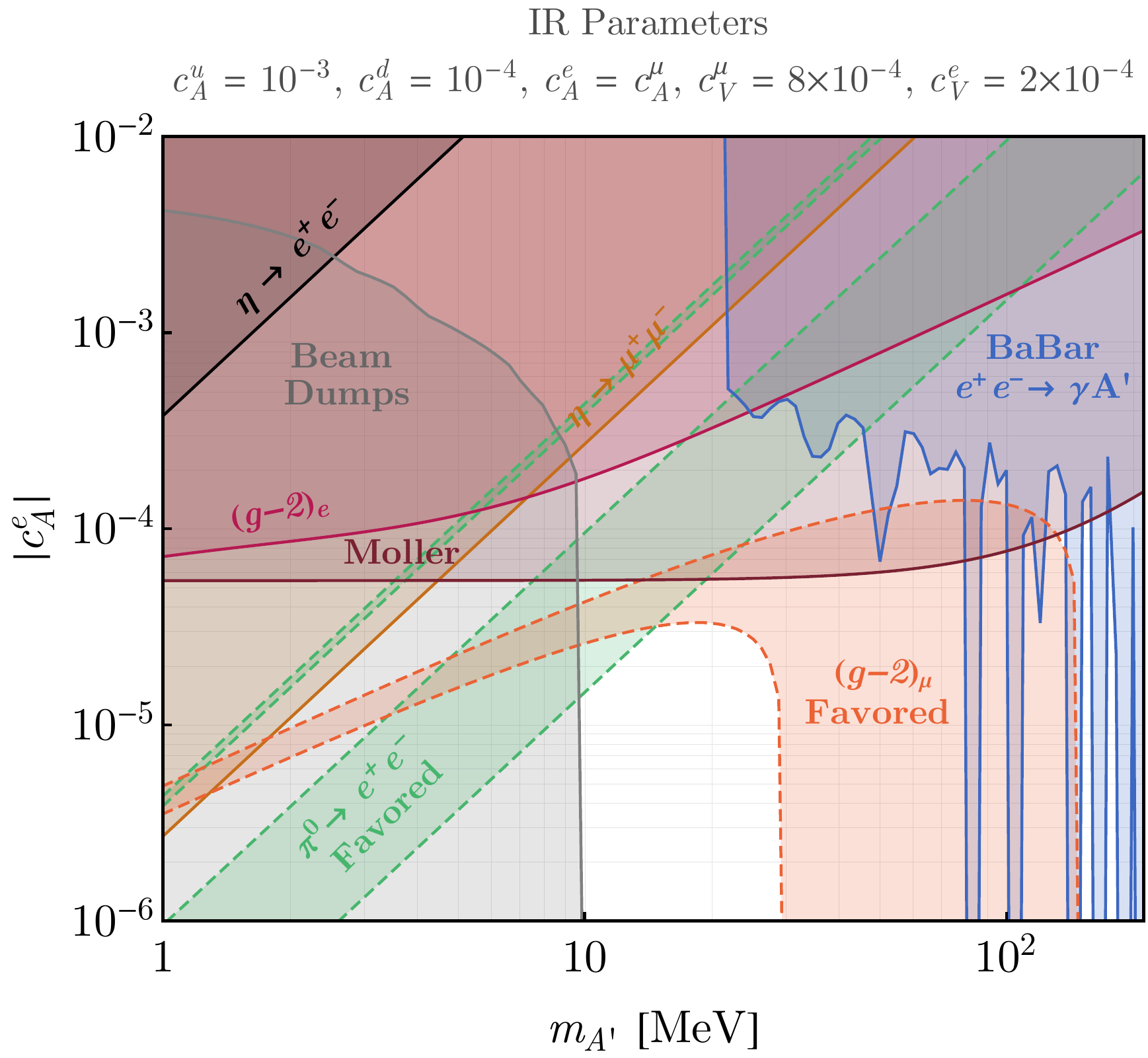} 
\caption{Parameter space for $c_A^e$ as a function of $m_{A'}$, assuming $c_A^u = 10^{-3}$ and $c_A^d = 10^{-4}$. The left panel assumes all other couplings vanish, while the right panel assumes $c_V^e =  2 \times 10^{-4}$, $c_V^\mu = 8 \times 10^{-4}$, and $c_A^\mu = c_A^e$. Two regions are compatible with $\pi^0 \to e^+ e^-$, but one of these is ruled out by $\eta \to \mu^+ \mu^-$. Note that specifying a value for $c_V^e$ changes the shape of \emph{e.g.}\ beam dump constraint curves for $c_A^e$ in the right panel.} 
\label{fig:NoVectors} 
\end{figure}

\begin{figure}[t!] 
\hspace*{-0.75cm}
\includegraphics[width=0.503\textwidth]{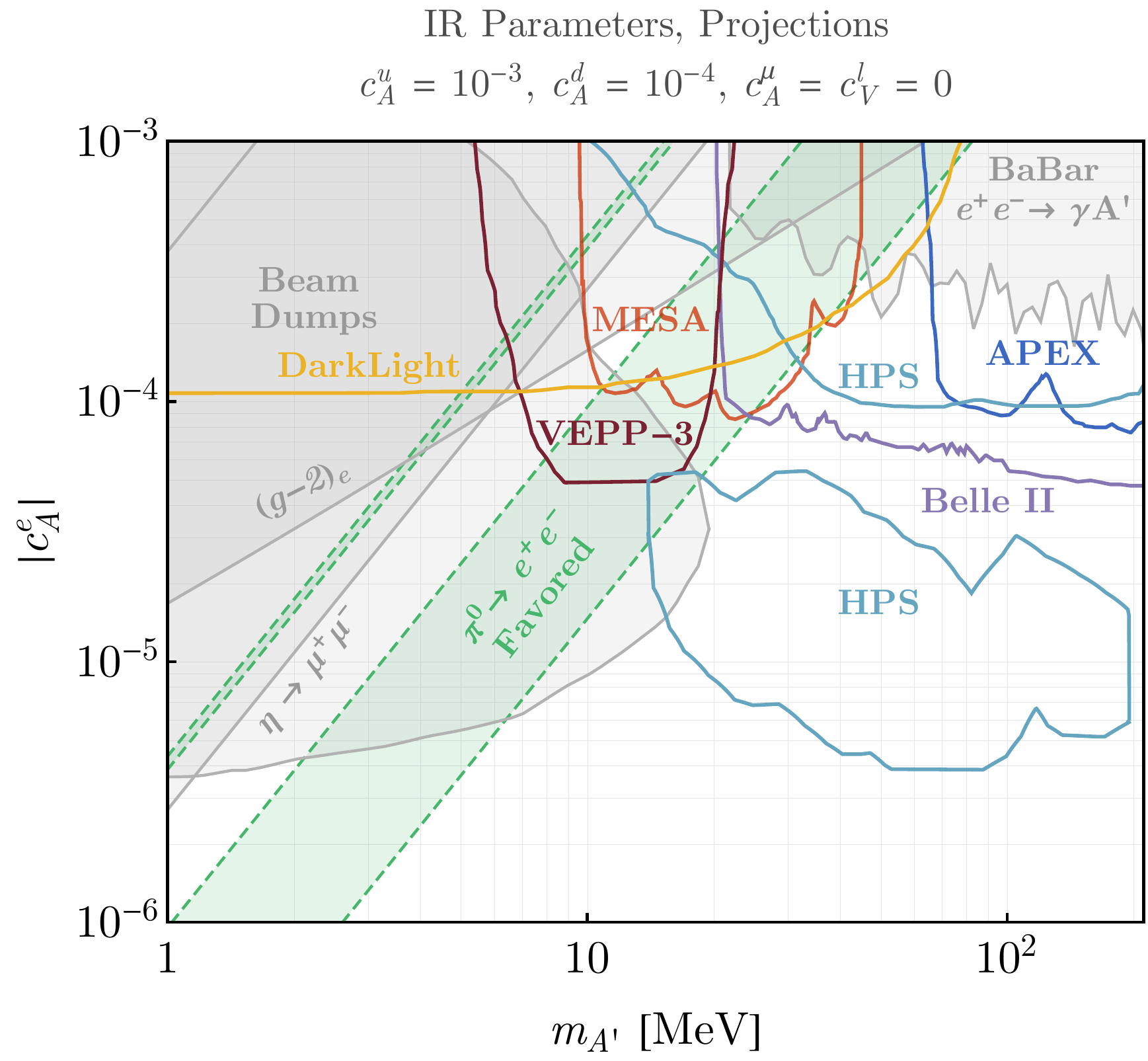} \qquad \hspace*{-0.3cm}
\includegraphics[width=0.499\textwidth]{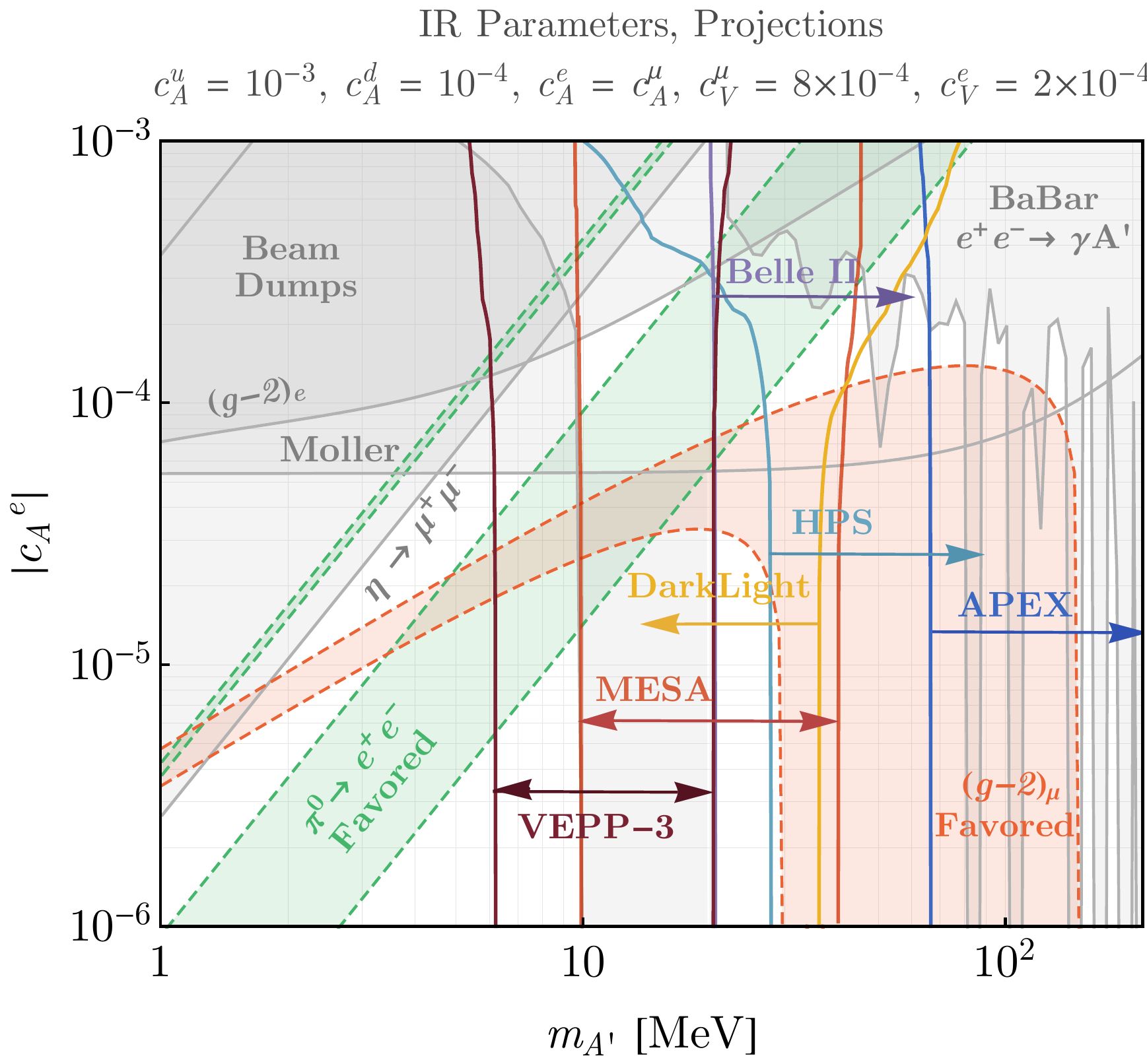} 
\caption{Projected constraints from various upcoming experiments, corresponding to the parameter space points shown in figure \ref{fig:NoVectors}. } 
\label{fig:Projected} 
\end{figure}

As a first pass through axially-coupled parameter space, consider the plots shown in figure \ref{fig:NoVectors}. Without regard to any constraints imposed by UV physics, we set all couplings in Eq.\ (\ref{eq:AprimeLag}) to zero except $c_A^{u,d,e,\mu}$ and $c_V^{e,\mu}$. These choices are motivated by the following observations, summarizing the discussion above:
\begin{itemize}
\item The decay $\pi^0 \to e^+ e^-$ prefers a nonzero value for the combination $c_A^e (c_A^u - c_A^d)$ (see \cite{Kahn:2007ru} and appendix \ref{app:Pseudoscalars}).\footnote{As we show in appendix \ref{app:Pseudoscalars}, since the $A'$ contribution interferes at tree-level with the SM loop contribution, there may be two solutions for the couplings and hence two disjoint preferred regions.}
\item The positive deviation of the measured $(g-2)_\mu$ compared to the SM prediction prefers $c_V^\mu \gg c_A^\mu$.
\item As noted in \cite{Fayet:2007ua}, the most constraining limits on $c_A^e$ come from cesium atomic parity violation measurements, but these vanish if $c_V^{u,d} = 0$. Likewise, the constraints on $A'$ production from $\pi^0 \to \gamma A'$, arising from experiments like NA48/2 \cite{Batley:2015lha}, also effectively vanish in this limit.
\end{itemize}
The left panel assumes all couplings other than $c_A^{u,d,e}$ vanish, while the right panel assumes nonzero vector and axial couplings for both $e$ and $\mu$. In the left panel, where $c_V^\mu = c_A^\mu = 0$, the $A'$ does not contribute to $(g-2)_\mu$ and one must assume that some other new physics contribution is responsible for the measured deviation from the SM value. Interestingly, the right panel shows a region ostensibly compatible with both the $\pi^0$ and $(g-2)_\mu$ anomalies, with couplings and $A'$ mass of the same order as the 17 MeV vector which was proposed to explain the recent $^8{\rm Be}$ anomaly. figure~\ref{fig:Projected} shows the landscape of projected constraints for both parameter space points. The compatibility region around 17 MeV can be probed by several upcoming experiments including VEPP-3 \cite{wojtsekhowski:2012zq}, DarkLight \cite{balewski:2014pxa,balewski:2013oza}, and MESA \cite{Aulenbacher:2013xla}, with additional parameter space covered by Belle II \cite{belle-ii-talk:2013}, HPS \cite{Celentano:2014wya}, and APEX \cite{abrahamyan:2011gv}.

It is worth pointing out that several experiments, including all the ones for which projected constraints are shown, only measure $\sqrt{(c_V^e)^2 + (c_A^e)^2}$ and do not independently constrain the axial couplings. Consequently, specifying a value for $c_V^e$ in parameter space can strongly affect the shape of the reach curve for $c_A^e$; if that value of $c_V^e$ is probed with $c_A^e = 0$, then it constrains \emph{all} values of $c_A^e$ at fixed $c_V^e$. Thus, many of the projections in figure~\ref{fig:Projected}, right (and the beam dump exclusion curves in figure~\ref{fig:NoVectors}, right, which depend on the same combination of parameters) extend all the way down to $c_A^e = 0$. Note that this effectively restricts the reach for beam dump experiments at high mass, because sufficiently large vector couplings force the $A'$ to decay before entering the detector region.

However, as we will see in section \ref{sec:Survey}, these choices of IR-motivated couplings are in strong tension with UV constraints. In particular, ensuring nonzero quark axial couplings with zero (or very small) quark vector couplings requires careful tuning in parameter space. UV completions of this model generically require relations among the parameters, either through gauge invariance of the SM Yukawa terms, or through the $\UD$ symmetry-breaking pattern which relates the $A'$ mass to the new heavy fermion masses. We will show that, absent some rather severe fine-tuning, the parameter space compatible with the $\pi^0$ anomaly is robustly excluded.

%


                                                               \section{Models with Gauge Invariant Yukawa Couplings}
                                                                                   \label{sec:GaugeInvtYukawas}


In this section we establish our framework and notation for models of axial forces whose charge assignments allow gauge-invariant Yukawa couplings for 
the SM fermions in the presence of single (section~\ref{sec:1HDM}) or multiple (section~\ref{sec:2HDM}) Higgs doublets. 
We begin by extending the SM to include a new local $\UD$ group with gauge boson $A'$ and gauge coupling $g_D$. 
In four-component notation, each SM fermion $f$ 
is represented as a Dirac spinor $f \equiv ( f_L, {f^{c}}^\dagger)$ where $f_L$ is a two-component Weyl spinor transforming as an $SU(2)_L$ doublet and $f^c$ is its 
Dirac partner identified with the corresponding singlet Weyl field; in our convention, all Weyl spinors are left-handed. Throughout, we will represent 
quarks as $u = (u_L, {u^c}^\dagger)$,  $d = (d_L, {d^c}^\dagger)$ and leptons as $e = (e_L, {e^c}^\dagger)$ and  $\nu = (\nu_L, 0),$ where we have suppressed flavor indices
and omitted right-handed neutrinos without loss of generality.

 Defining the vector  and axial-vector currents  
\be
\label{eq:VectorAxialDef}
~~J_V^\mu \equiv  \bar f \gamma^\mu f =  f_L^\dagger\bar \sigma^\mu f_L  -  {f^c}^\dagger \bar \sigma^\mu f^c , ~~~~~~  
J_A^\mu \equiv  \bar f \gamma^\mu \gamma^5 f =  f_L^\dagger\bar \sigma^\mu f_L  +  {f^c}^\dagger \bar \sigma^\mu f^c,
\ee
for each $f$ we have vector ($c^f_V$) and axial-vector ($c^f_A$) couplings to the $A'$,
\be \label{eq:ax-vec-coupling}
c^f_V \equiv \frac{1}{2}g _D ( q_{f_L} - q_{f^c}),~~~~ c^f_A \equiv \frac{1}{2}  g_D (q_{f_L} + q_{f^c}),
\ee
where $q_{f_L, f^c}$ are the charges of the appropriate Weyl spinor under $\UD$.


		\subsection{Axial Cancellation for Single Higgs Models}
		                 	\label{sec:1HDM}


We begin with models containing a single Higgs doublet, demonstrating that such models do {\em not} lead to large axial-vector couplings to SM fermions
when charges are assigned such that the SM Yukawa interactions are permitted by $\UD$ gauge invariance.

			\subsubsection{Generic Properties  of an Axial $U(1)$}
			                      \label{sec:genprop}

To illustrate the general features of an axially-coupled $U(1)$ gauge extension, consider the minimal Yukawa sector with 
a single Higgs doublet 
\be
\label{eq:1HDMYukawa}
{\cal L}_{y \cal, {\rm SM}} =  y_u H Q u^c  +   y_d H^\dagger Q d^c   + y_e H^\dagger L e^c  + {h.c.},
\ee
where  $y_{u,d,e}$ are $3 \times 3$ Yukawa matrices, $H$ is the SM Higgs doublet, and $Q = (u_L, d_L)$ and  $L = (\nu_L, e_L)$ are respectively the quark and lepton doublets. By
stipulation, the Yukawa couplings are gauge invariant under $\UD$, so based on Eq.~(\ref{eq:ax-vec-coupling}) we have 
\be
  q_{f_L} + q_{f^c} \neq 0~~ \to ~~   q_H = -(q_{f_L} + q_{f^c}) \neq 0,
\ee
where $q_f$ represents the $\UD$ charge of $f$ and the Higgs doublet  carries a nonzero $\UD$ charge $q_H$. These charge assignments are equivalent to $U(1)_{\alpha B + \beta L + \gamma Y}$, a linear combination of the two accidental symmetries of the SM in the IR, $B$ and $L$, and hypercharge $Y$. As is well-known, taking $\alpha = -\beta$ gives the  anomaly-free group $B-L + xY$.

Prior to EWSB, the axial couplings are
 \be \label{eq:nomixcouplings}
 \hspace{2cm }c_{A}^d = c_{A}^e = -c_{A}^u  =  \frac{1}{2} g_D q_H ~~ \to  ~~  c_{A}^f =  - g_D q_H  T^3_f~~~~~~~~     ({\rm before ~ EWSB}),
 \ee
 where $T^3$ is the diagonal generator of $SU(2)_L$ and $T^3_f$ is its eigenvalue for a given fermion.  After EWSB, the Higgs VEV contributes to $\UD$ breaking, and the neutral gauge boson mass terms can
 be written
\be \label{eq:massterms}
{\cal L } &=& 
\frac{1}{2}(\hat Z_\mu~~ \hat A^\prime_\mu) 
\left( 
{\begin{array}{cc}
   \hat{m}_{Z}^2 &
  - g_D q_H v \hat{m}_Z       
   \\
  - g_D q_H  v \hat{m}_Z    
   &  ~~~ g^2_D q_H^2 v^2      + \hat{m}_{A^\prime}^2
\end{array} } 
   \right)
   \left(
\begin{array}{c} 
\hat{Z}_\mu\\
\hat{A}^\prime_\mu\\
\end{array}
\right),
\ee
where $\hat{m}_Z = g v/2 c_W$ is the $Z$ boson mass in the SM, $g$ is the $SU(2)_L$ gauge coupling, $c_W \equiv \cos \theta_W$, $\theta_W$ is the weak mixing angle, and for future convenience 
we define $s_W \equiv \sin \theta_W$ and $t_W \equiv \tan \theta_W$.  Throughout this work, we define gauge boson interaction eigenstates as $\hat Z$, $\hat A^\prime$. In Eq.~({\ref{eq:massterms}}) we have also included
a $\hat{m}_{A^\prime}^2$ contribution in the lower-right entry, representing an additional hidden-sector source for the $A^\prime$ mass, \emph{e.g.} from dark Higgses which are SM singlets. The phenomenology of a ``dark $Z$'' with a general mass mixing matrix has been investigated in \cite{Davoudiasl:2012ag,Davoudiasl:2012qa,Davoudiasl:2014kua,Lee:2014lga,Lee:2016ief}. We diagonalize to the mass eigenbasis  $( Z_\mu, 
{ A}^\prime_\mu)$,
\be \label{eq:massmatrix-nokin}
   \left(
\begin{array}{c} 
{\hat Z}_\mu\\
{\hat A}^\prime_\mu\\
\end{array}
\right) =
\left( 
{\begin{array}{cc}
\cos \eta &
\sin \eta  \\ 
-\sin \eta  
   &
 \cos \eta       
\end{array} } 
   \right)
      \left(
\begin{array}{c} 
{Z}_\mu\\
{A}^\prime_\mu\\
\end{array}
\right),
\ee
where the mixing angle satisfies
\be \label{eq:taneta}
\tan 2 \eta  =  \frac{ - 2 g_D q_H  v \hat{m}_Z         }{  ( g_D^2 q_H^2 v^2 + \hat{m}_{A^\prime}^2     )    -\hat{m}_{Z}^2}.
\ee
In the $g_D q_H \ll1$, $ \hat{m}_{A^\prime}  \ll \hat{m}_Z$ limit, we have
\be
\label{eq:tanetaexpand}
\sin \eta  \simeq  \frac{g_D q_H v}{\hat{m}_Z} \left( 1 -  \frac{g^2_D q^2_H v^2}{\hat{m}_Z^2}  - \frac{\hat{m}^2_{A^\prime}}{\hat{m}_Z^2} \right),
\ee
and we can write the induced coupling to fermions as 
\be
\hat{Z}_\mu J_{\rm NC}^\mu  = (  \cos \eta  \, Z_\mu +   \sin \eta \, A^{\prime}_\mu ) J_{\rm NC}^\mu,
\ee
with $\sin \eta \simeq g_D q_H v / \hat{m}_Z $. Thus, the physical $A'$ inherits a coupling proportional to the weak neutral current through mass mixing, which can be written
\be 
\sin \eta \, A^{\prime }_\mu  J_{\rm NC}^\mu \simeq \frac{g_D q_H g v}{  \hat{m}_Z    c_W }    A^{\prime }_\mu  \,    \sum_i    (T^3  - Q_{\psi_i} s^2_W) \psi_i^\dagger \bar \sigma^\mu \psi_i,
\ee
where the sum is over all 2-component SM fermions  $\psi_i = Q, L, u^c, d^c, e^c $. 
The axial coupling from the SM neutral current is $c^{f}_{A,\rm SM} =   \frac{g}{2 c_W} T^3_f$, so the axial coupling induced by mixing with the $Z$ is   
\be
{ \Delta c_{A}^f} = \sin \eta \,  c^{f}_{A, \rm SM} \simeq  \frac{ g_D q_H v}{ \hat{m}_Z}   \frac{g}{2c_W} T^3_f   
  =  + g_D q_H   T^3_f.
\ee
Adding this contribution to the UV axial coupling from Eq.~(\ref{eq:nomixcouplings}), we get
\be
\hspace{3cm}            c_A^f \to  c_A^f +\Delta { c_{A}^f} = 0   + \mathcal{O}(g_D^3)  + \mathcal{O}(\hat{m}^2_{\apr}/\hat{m}^2_Z)   ~~~~~~~~~~~~~~~ ({\rm after~EWSB}),
\ee 
so the axial coupling cancels up to small corrections of order $g_D^3$ and $\hat{m}^2_{\apr}/\hat{m}^2_Z$ in Eq.~(\ref{eq:tanetaexpand}). 

The cancellation in Eq.~(\ref{eq:axial-afterEWSB-1HDM}) is generic for any $U(1)$ extension of the SM in which there is a
single Higgs doublet whose Yukawa couplings respect $\UD$ gauge invariance; it does not depend on any additional field content which might accompany such an extension (\emph{e.g.} to cancel anomalies).
On physical grounds, this cancellation occurs for $\hat m_{A^\prime} \ll \hat m_Z$ because this limit introduces a zero eigenvalue for the mass matrix and approximately restores the $U(1)_D$ symmetry even though SM fermions have acquired mass from the Higgs doublet after EWSB, so their axial current is no longer invariant under $U(1)_D$. As we will see below in section\ref{sec:2HDM}, this cancellation is not guaranteed in  a 
two Higgs doublet model where the $\hat m_{A^\prime} \ll \hat m_Z$ limit does not necessarily restore the $U(1)_D$ symmetry; the additional electroweak doublet can now
give mass to $A^\prime$ even in this regime, so its axial couplings need not vanish. 

 Furthermore, it is intriguing that the axial couplings in Eq.~(\ref{eq:nomixcouplings})  can be sizable in the very early universe, 
only to vanish in Eq.~(\ref{eq:axial-afterEWSB-1HDM}) after EWSB; they serve as an order parameter for the electroweak phase transition. Understanding the  cosmological implications of such a  cancellation is beyond the scope of this paper, but we note that this mechanism may allow otherwise dangerous axial couplings to be sizable in the early universe, only
to cancel at late times when the electroweak symmetry is broken.

 			\subsubsection{Kinetic Mixing}
			\label{sec:1HDM-kinmix}

 Our gauge extension also allows a renormalizable kinetic mixing between $U(1)_Y$ and $U(1)_D$,
 \be
 {\cal L } \supset -\frac{1}{4} {\hat B}^{\mu \nu}  {\hat B}_{\mu \nu}      +  \frac{\epsilon}{2 c_W}  \hat{B}^{\mu \nu} \hat{F}^\prime_{\mu \nu} 
 -\frac{1}{4} \hat{F}^{\prime \, \mu \nu}\hat{F}^\prime_{\mu \nu},
 \ee
 where $\epsilon \ll 1$ is the kinetic mixing parameter, and the fields are written 
in terms of the interaction eigenstates $\hat B$ and $\hat A$. We can diagonalize away the kinetic mixing by shifting the hypercharge field 
 \be\label{eq:bshift}
 \hat B_{\mu } \to   \hat  B_{\mu } + \frac{\epsilon}{c_W}  \hat  A^\prime_\mu,
 \ee
 which eliminates the off-diagonal $\hat B^{\mu \nu} \hat F^\prime_{\mu \nu}$ term and rescales 
 $\hat F^\prime_{\mu\nu} \to (1 + \epsilon^2/c^2_W)\hat F^\prime_{\mu\nu}$ by a negligible, ${\cal O}(\epsilon^2)$ amount. In terms of the
 IR interaction eigenstates $\hat A, \hat Z$, and $\hat \apr$, the shift in Eq.~(\ref{eq:bshift}) is equivalent to
 \be
\hat A_\mu \to \hat A_\mu  + \epsilon \hat \apr_\mu,~~~~\hat Z_\mu \to \hat Z_\mu  - \epsilon t_W \hat \apr_\mu,~~~~~ \apr_\mu \to \apr_\mu, 
 \ee
 so $\apr$ acquires ${\cal O}(\epsilon)$ couplings to SM fermions of the form 
 \be\label{eq:currents}
 e \hat A_\mu J^\mu_{\rm EM} \to  e(\hat A_\mu + \epsilon \hat  A^\prime_\mu) J^\mu_{\rm EM},~~ g_Z \hat Z_\mu J^\mu_{\rm NC} \to  g_Z( \hat  Z_\mu - \epsilon t_W  \hat  A^\prime_\mu) J^\mu_{\rm NC},  
 \ee
 where $g_Z \equiv g/c_W$ is the 
SM $Z$ coupling and $J_{\rm EM}$ and $J_{\rm NC}$ are respectively the electromagnetic and neutral currents. 
This shift in  Eq.~(\ref{eq:bshift}) also induces a correction to the mass terms 
\be \label{eq:massterms-withkin}
\nonumber \\
{\cal L } &\supset& 
\frac{1}{2}(\hat Z_\mu~~ \hat A^\prime_\mu) 
\left( 
{\begin{array}{cc}
   \hat{m}_{Z}^2 &
  - \left(g_D q_H    +  \frac{1}{2} \epsilon\, t_W  g_Z        \right) v \hat{m}_Z       
   \\
  - \left(g_D q_H    +  \frac{1}{2} \epsilon \, t_W g_Z         \right) v \hat{m}_Z  
   &~~~~  \left(g_D q_H + \frac{1}{2} \epsilon t_W g_Z \right)^2 v^2      + \hat{m}_{A^\prime}^2
\end{array} } 
   \right)
   \left(
\begin{array}{c} 
\hat{Z}_\mu\\
\hat{A}^\prime_\mu\\
\end{array}
\right),
\ee
\noindent which is equivalent to shifting $g_D q_H \to g_D q_H+ \epsilon t_W g_Z /2$ in Eq.~(\ref{eq:massterms}).\footnote{This  agrees with the 
results of  \cite{Lee:2016ief} which considered the  ``mini-force" model with the $U(1)_{B-L + x Y}$ group. We recover their results 
 for $q_H = x/2$, since the mass mixing phenomenology is uniquely determined by the Higgs charge under the new $U(1)$.}
 This matrix has the same structure
 as Eq.~(\ref{eq:massterms-withkin}), so again diagonalizing with an
 orthogonal rotation, the mixing angle $\zeta$ satisfies 
\be \label{eq:taneta-epsilon}
\tan 2 \zeta  =  \frac{ - 2\left( g_D q_H      + \frac{1}{2} \epsilon t_W g_Z   \right) v \hat{m}_Z         }{  \left( g_D q_H + \frac{1}{2}\epsilon t_W g_Z \right)^2 v^2 + \hat{m}_{A^\prime}^2  -\hat{m}_{Z}^2}.
\ee
 Note that in the pure kinetic mixing limit, $g_D q_H \to 0$, with $\hat m_{\apr} \ll \hat m_Z$ we have $\zeta \ll 1$, so this expression yields
 \be
 \sin \zeta =
\frac{ - \left(  \frac{1}{2} \epsilon t_W g_Z   \right) v \hat{m}_Z         }{  \left( \frac{1}{2}\epsilon t_W g_Z \right)^2 v^2 + \hat{m}_{A^\prime}^2  -\hat{m}_{Z}^2}
\simeq \epsilon t_W g_Z    \left( 1 - \epsilon^2 t^2_W -   \frac{  \hat{m}_{A^\prime}^2}{ \hat{m}_{Z}^2}   \right),
 \ee
 so writing the interaction eigenstates in terms of the mass eigenstates $Z, A$ gives  
 \be
 \hat Z_\mu = \cos \zeta Z_\mu + \sin\zeta A^\prime_\mu,~~~~   \hat A^\prime_\mu = -\sin \zeta Z_\mu + \cos \zeta A^\prime_\mu.
 \ee
Thus, in the mass basis, the neutral current interaction in Eq.~(\ref{eq:currents}) becomes 
\be
g_Z ( \hat  Z_\mu - \epsilon t_W  \hat  A^\prime_\mu) J^\mu_{\rm NC} 
&\simeq&
g_Z \left( Z_\mu + \epsilon t_W A^\prime_\mu - \epsilon t_W A^\prime_\mu   \right) J^\mu_{\rm NC} + {\cal O}(\epsilon^3) + {\cal O}(\epsilon \hat m_{\apr}^2/\hat m_Z^2),
\ee
where we recover the familiar cancellation of  the $\apr$ neutral current coupling to ${\cal O}(\epsilon)$; the leading $\apr-A$ interaction in Eq.~(\ref{eq:currents}) survives the $\zeta$
 rotation, so a light, kinetically mixed gauge boson is properly a ``dark photon" and not a ``dark $Z$" boson. 
 We see that this well known feature of $\apr-Z$ mixing has the same origin as the cancellation of axial couplings presented in section \ref{sec:genprop}, where
 we considered the $g_Dq_H \ne 0, \epsilon = 0$ regime.
 
  Although the minimal examples considered in this section yield
 only suppressed axial couplings, they highlight the generic limits of axial $U(1)$ models with gauge-invariant Yukawa interactions and a single Higgs boson. 
 Furthermore, the machinery and formalism developed in this section will prove useful below where we consider extended Higgs sector models 
 for which this cancellation no longer takes place and unsuppressed axial couplings are generically present.


	\subsection{Scenarios with Two Higgs Doublets (2HDM)}
	\label{sec:2HDM}

We now construct a model which \emph{does} result in unsuppressed axial couplings below the electroweak scale. The full model contains several ingredients which control the vector and axial couplings of SM fermions, some in a generation-dependent way. For pedagogical purposes we will build up the model one ingredient at a time, with the full $U(1)_D$ group presented in section~\ref{sec:fullmodel}.

		\subsubsection{Generic Properties }

\label{sec:Model}

Consider now a Type-II two-Higgs-doublet model (2HDM) where $H_u$, $H_d$ and right-handed SM fermions are charged under a new $\URH$ with gauge coupling $g_D$.\footnote{One could also consider a ``flipped'' 2HDM, where the same Higgs doublet provides masses for the up-type quarks
together with the charged leptons.}
Charging the right-handed SM fermions is a specific choice, but one
which does not lead to an essential loss of generality.\footnote{A similar charge assignment was considered in \cite{Appelquist:2002mw}.} As in the single Higgs doublet model, the fermion $U(1)_D$ charges are related to the Higgs charges: $q_{u^c} = -q_{H_u}$ and $q_{d^c} = q_{e^c} = -q_{H_d}$. 
With these charge assignments, the SM Yukawa terms are invariant under $\URH$:
\be
{\cal L}_{\cal Y, {\rm 2HDM}} =  y_u H_u Q u^c  +   y_d H_d Q d^c   + y_e H_d L e^c + {\rm h.c.}
\ee
This group is anomalous under the SM, so we add anomalons ${\cal U}/{\cal U}^c$, ${\cal D}/{\cal D}^c$, and ${\cal E}/{\cal E}^c$ which are vector-like under the SM and  chiral under $\UD$ to cancel 
gauge anomalies.  Furthermore, all anomalons considered here are electroweak singlets, so there is only minimal impact on precision electroweak observables.
Two dark Higgses $H'_u$ and $H'_d$ are required in order to give masses to the anomalons:
\be
\label{eq:2HDMAnomalon}
{\cal L} = {\cal L}_{{\cal Y}, {\rm 2HDM}} +   y_{\cal U} H'_{u} \, {\cal U}{\cal U}^c + y_{\cal D} H'_{d} {\cal D}{\cal D}^c    +   y_{\cal E} H'_{d} {\cal E}{\cal E}^c + {\rm h.c.}
\ee
The field content and charge assignments for this setup (shown in table \ref{tab:U1RCharges}) are chosen to guarantee anomaly cancellation without contributions from additional dark sector fields. If $q_{H_u} = -q_{H_d}$, we recover the single Higgs doublet phenomenology discussed in section \ref{sec:1HDM}, in which axial couplings cancel at low energies to leading order.

Before moving on to the consequences for axial and vector couplings, we note a few relevant features of this model:
\begin{itemize}
\item Dangerous trilinear terms, which would generate tadpoles when the various Higgses get vacuum expectation values (VEVs), are forbidden 
from the Higgs potential by gauge invariance.
\item The dark Higgses are SM singlets and so do not contribute to electroweak symmetry breaking. On the other hand, all four Higgses contribute to the mass
of the $A'$.
\item Mass mixing of the form $m^U_{ij} {\cal U}_i u_j^c + m^D_{ij} {\cal D}_i d_j^c + m^E_{ij} {\cal E}_i e_j^c$ is allowed by gauge invariance and permits the anomalons to decay into SM fermions. 
These mixings (which are not necessarily proportional to the SM Yukawas) are technically natural, and can be small enough to evade bounds on the unitarity of 
the CKM matrix and flavor-changing neutral currents (FCNCs), while still allowing for prompt decays on collider scales, since the latter only requires one mass insertion while the former requires two. Alternatively, one could impose minimal flavor violation (MFV) on the mass mixing matrices, in which case FCNCs would be absent at leading order.

\end{itemize} 
The colored anomalon masses are typically bounded by LHC searches for
fourth-generation quarks, which require their masses to be $\gtrsim 1 \ \TeV$  \cite{Khachatryan:2015oba}.
This collider constraint on the UV theory has interesting IR implications, because both the $A'$ and anomalon masses receive contributions from
the dark Higgs VEV $v'$ and perturbativity requires
the Yukawa interactions of Eq.~(\ref{eq:2HDMAnomalon}), generically denoted by $y_\psi$, to 
satisfy $y_\psi \lesssim 4\pi$.\footnote{A more stringent constraint would require that both the anomalon Yukawas and the hypercharge
coupling remain free from Landau poles below the Planck scale.  We content ourselves with the requirement that the theory be self-consistent at the TeV scale.} Since the dark Higgs will also contribute to the $A'$ mass, we have $m_{A'} \gtrsim g_D q_H v'$, leading to the constraint
\be
\label{eq:anomalonconstraint}
m_{A'} \gtrsim  80 \ {\rm MeV} \times \left(\frac{g_D q_H}{10^{-3}}\right) \times \left(\frac{4\pi}{y_\psi}\right),
\ee
implying that very light axially-coupled $A'$s are required by LHC searches to be also very weakly coupled. Note that we have restored the factor of $q_H$ in this bound compared to the rough estimate (\ref{eq:anomalonconstraintrough}), and that if multiple Higgses are present, this bound will be correspondingly strengthened.

\subsubsection{Axial Couplings from $\URH$}

\begin{table}[t] 
\hspace{-6cm}
\begin{center}
\vline
\begin{tabular*}{0.65\textwidth}{@{\extracolsep{\fill}}lclclclclc|c|}
\hline
\\[-7pt] 
$\quad$                               &{\bf Field}&                                         $SU(3)_c$     &&     $SU(2)_L$            &&         $U(1)_Y$   &&         $\URH$    \\[2pt]
\hline
\\[-6pt]
$\quad$                               & $H_u$  &                                         ~~~${\bf  1}$         &&                       ~~~${\bf 2}$           &&      ~~$ +\frac{1}{2} $  &&         $+q_{H_u}$    \\[2pt]
$\quad$                               & $H_d$  &                                         ~~~${\bf  1}$         &&                       ~~~${\bf 2}$           &&      ~~$ -\frac{1}{2} $  &&         $+q_{H_d}$    \\[2pt]

\\[-6pt]
\hline
\\[-6pt]
$\quad$                               & $u^c$  &                                     ~~~${\bf \overline 3}$     &&           ~~~${\bf 1}$            &&      ~~$ -\frac{2}{3} $  &&         $-q_{H_u}  $     \\[2pt]
$\quad$                               & $d^c$  &                                     ~~~${\bf \overline 3}$     &&           ~~~${\bf 1}$            &&      ~~$ +\frac{1}{3} $  &&        $-q_{H_d}      $    \\[2pt]
$\quad$                               & $e^c$  &                                     ~~~${\bf  1}$     &&                           ~~~${\bf 1}$           &&      ~~$+1 $  &&         $-q_{H_d}$    \\[2pt]
\\[-7pt] 
\hline\hline
\\[-7pt] 
%
$\quad$                               & ${\cal U}$  &                                         ~~~${\bf  3}$     &&           ~~~${\bf 1}$            &&      ~~$ +\frac{2}{3} $  &&                          $+q_{H_u}$    \\[2pt]
$\quad$                               & ~${\cal U}^c$  &                                     ~~~${\bf \overline 3}$     &&           ~~~${\bf 1}$            &&      ~~$ -\frac{2}{3} $  &&         $~~~0$    \\[2pt]
$\quad$                               & ${\cal D}$  &                                     ~~~${\bf  3}$     &&           ~~~${\bf 1}$            &&      ~~$ -\frac{1}{3} $  &&         $+q_{H_d}$    \\[2pt]
$\quad$                               & ~${\cal D}^c$  &                                     ~~~${\bf \overline 3}$     &&           ~~~${\bf 1}$            &&      ~~$ +\frac{1}{3} $  &&         $~~~0$    \\[2pt]
$\quad$                               & ${\cal E}$  &                                      ~~~${\bf  1}$     &&                           ~~~${\bf 1}$           &&      ~~$ -1 $  &&         $+q_{H_d}$    \\[2pt]
$\quad$                               & ~${\cal E}^c$  &                                      ~~~${\bf  1}$     &&                           ~~~${\bf 1}$           &&      ~~$ +1 $  &&         $~~~0$    \\[2pt]
\\[-6pt]
 \hline
$\quad$                               &~ $H^\prime_u$  &                                    ~~~${\bf  1}$         &&                       ~~~${\bf 1}$           &&      ~~$ ~~0 $  &&         $-q_{H_u}$    \\[2pt]
$\quad$                               &~ $H^\prime_d$  &                                    ~~~${\bf  1}$         &&                       ~~~${\bf 1}$           &&      ~~$ ~~0 $  &&         $-q_{H_d}$    \\[2pt]
\hline

 	\end{tabular*}
\hspace{-0.2cm}	\vline
\caption{$\URH$ charge assignments for for a type-II 2HDM scenario. The SM fields $Q$ and $L$ are neutral under $\URH$. Three generations of fermions are understood.}
\label{tab:U1RCharges}
\end{center}
\end{table}

A key feature of this model is that the same Higgs doublet couples to all three generations of each type of fermion, implying that the $U(1)_{RH}$ axial couplings are the same for each generation. In this setup, there are two independent axial couplings, parameterized by the two Higgs charges $q_{H_u}$ and $q_{H_d}$.  As in the single Higgs scenario above, using the definitions in Eq.~(\ref{eq:VectorAxialDef}) the axial couplings before EWSB are simply 
related to the Higgs charges under $\URH$:
\be \label{eq:axial2HDM_UV}
~~~~~~~~~~~~~~~~ c_{A}^{u} = -\frac{1}{2} g_D q_{H_u},~~~~~ c_{A}^{d} = c_{A}^{e}  =   -\frac{1}{2} g_D q_{H_d} ~~~~~~~~~~~ {\rm (before ~EWSB)}.
\ee
After EWSB, $\langle H_u \rangle = \frac{1}{\sqrt{2}}(0,v_u)$ and $\langle H_d \rangle = \frac{1}{\sqrt{2}}(v_d,0)$, with $v^2 = v_u^2 + v_d^2  = (246\, \GeV)^2$, so 
 the neutral gauge boson mass matrix is 
\be \label{eq:massterms2HDM}
\frac{1}{2}(\hat Z_\mu~~ \hat A^\prime_\mu) 
\left( 
{\begin{array}{cc}
   \hat m_Z^2 &
  - g_D(q_{H_u} v_u^2 - q_{H_d} v_d^2)  \hat m_Z/v       
   \\
  - g_D(q_{H_u} v_u^2 - q_{H_d} v_d^2) \hat m_Z/v       
   &  g^2_D (q_{H_u}^2 v_u^2   + q_{H_d}^2 v_d^2)      + \hat m_{\apr}^2
\end{array} } 
   \right)
   \left(
\begin{array}{c} 
\hat Z_\mu\\
\hat A^\prime_\mu\\
\end{array}
\right), \\ \nonumber
\ee
where for now we have neglected the effects of $U(1)_Y-U(1)_{RH}$ kinetic mixing.
This matrix is diagonalized with a rotation angle $\theta_D$, which satisfies 
\be
\tan 2 \theta_D  =  \frac{ - 2 g_D (q_{H_u} v_u^2 - q_{H_d} v_d^2)   \hat{m}_Z /v         }{  g_D^2 ( q_{H_u}^2 v_u^2 + q_{H_d}^2 v_d^2) + \hat{m}_{A^\prime}^2        -\hat{m}_{Z}^2}.
\ee
In the $g_D \ll 1$, $\hat m_{\apr} \ll \hat m_Z$ limit, this can be written as  
\be
\sin \theta_D  \simeq  \theta_D  \simeq  \frac{   g_D (q_{H_u} v_u^2 - q_{H_d} v_d^2)       }{   \hat{m}_{Z} v}  =   \frac{   2 g_D (q_{H_u} v_u^2 - q_{H_d} v_d^2)       }{ g_Z v^2}   \equiv \frac{2 g_D }{g_Z} \tilde \theta_D,
\ee
where we have defined $\tilde \theta_D \equiv (q_{H_u} v_u^2 - q_{H_d} v_d^2)/v^2$ for future convenience. As in section~\ref{sec:1HDM}, induced $\apr$ neutral current interactions arise from 
$\hat Z - \hat \apr$ mixing after rotating into the mass basis:
\be 
  \hat Z_\mu J^\mu_{\rm NC}  =   (\cos \theta_D Z_\mu + \sin \theta_D \apr_\mu) J^\mu_{\rm NC} \simeq   \left( Z_\mu + \frac{2g_D}{g_Z} \tilde \theta_D \apr_\mu \right) J^\mu_{\rm NC}.
\ee
However, unlike the result in section~\ref{sec:1HDM}, the $\apr$ axial coupling with SM fermions does not cancel. Since the SM axial coupling between the $Z$ and fermion $f$ is
 $c_{A, \rm SM}^f = \frac{g_Z}{2} T^3_f$, the induced coupling to the $\apr$ is $\Delta c_A^f \simeq \theta_D c_{A, \rm SM}^f$; combining this shift with the UV contribution in Eq.~(\ref{eq:axial2HDM_UV})
 the $\apr$ axial couplings become
 \be
 c_{A}^u = -\frac{1}{2} g_D q_{H_u} + \frac{1}{2} g_D \tilde \theta_D,~~~~ c_{A}^d = c_{A}^e =  -\frac{1}{2} g_D q_{H_d} - \frac{1}{2} g_D \tilde \theta_D~~~({\rm after~EWSB}), 
 \ee
which are nonzero for generic values of $v_u$ and $v_d$. Note that for $q_{H_u} = -q_{H_d}$ the mixing parameter is  $\tilde \theta_D = q_{H_u}$  and
we recover the earlier cancellation from the single Higgs scenario in section~\ref{sec:1HDM}; in this regime, the two Higgs VEVs are aligned and there is only one source 
of EWSB (which only gives mass to the $Z$ boson) so $\hat m_{A^\prime} \ll \hat m_Z$ limit approximately restores $U(1)_{RH}$, which forbids axial 
coupling to massive fermions.
\subsubsection{Full $U(1)_D$}
\label{sec:fullmodel}

By itself, $\URH$ leads to family-universal couplings of the SM fermions.  However, the group $U(1)_{\mu - \tau}$ is anomaly-free with respect to both the SM and $\URH$,
and the small breaking implied by neutrino oscillations does not significantly impact the viable parameter space for the masses considered here
(in contrast to similar constructions involving the quarks).
Its inclusion in the full dark gauge group allows the consideration of generation-dependent vector couplings for SM 
leptons, although their axial couplings 
are still fixed by $\URH$, and thus remain generation-independent.
We are led to consider the dark gauge group
\be
U(1)_D \equiv U(1)_{RH + \kappa(L_\mu - L_\tau)}
\ee
where $\kappa$ is a real parameter characterizing the relative importance of $L_\mu - L_\tau$.\footnote{Other permutations of lepton number, $U(1)_{e-\mu}$ and $U(1)_{e - \tau}$, are also possible, but their inclusion is easily mimicked with
the ingredients already at hand, and so we omit them without loss of essential generality.}

\begin{table}[t!]
\begin{center}
\hspace{-1.cm}
\begin{tabular*}{\textwidth}{@{\extracolsep{\fill}}c|c|c p{6cm}}
\bf{SM lepton} & $e$ & $\mu,\tau$  \hspace{-6cm}\\
 \hline
 $c_V^\ell$ & $\frac{1}{2} g_D q_{H_d} - \epsilon e  + g_D \tilde{\theta}_D(-\frac{1}{2} + 2 s^2_W)$ & $g_D (\frac{1}{2} q_{H_d}\pm \kappa) - \epsilon e + g_D \tilde{\theta}_D(-\frac{1}{2} + 2 s^2_W)$ \\
 \hline 
 $c_A^\ell$ & $- \frac{1}{2} g_D q_{H_d} - \frac{1}{2} g_D \tilde{\theta}_D$ & $- \frac{1}{2}  g_D q_{H_d}  - \frac{1}{2} g_D \tilde{\theta}_D$ 
 	\end{tabular*}	
\caption{Axial and vector couplings for SM charged leptons in a 2HDM scenario. In the top-right entry, the plus sign applies to $c_V^\mu$ and the minus sign to $c_V^\tau$.}
\label{tab:VAChargesL}
\end{center}
\end{table}

\begin{table}[t!]
\begin{center}
\begin{tabular*}{0.9\textwidth}{@{\extracolsep{\fill}}c|c|c}
\bf{SM quark} & $u,c,t$ & $d,s,b$ \\
\hline
 $c_V^q$ & $\frac{1}{2} g_D q_{H_u} + \frac{2}{3}\epsilon e + g_D \tilde{\theta}_D(\frac{1}{2} - \frac{4}{3} s^2_W)$ & $  \frac{1}{2}  g_D q_{H_d} - \frac{1}{3}\epsilon e + g_D \tilde{\theta}_D(-\frac{1}{2} + \frac{2}{3} s^2_W)$\\
 \hline 
 $c_A^q$ & $- \frac{1}{2}  g_D q_{H_u} + \frac{1}{2}g_D \tilde{\theta}_D $ & $-\frac{1}{2}  g_D q_{H_d} - \frac{1}{2}g_D \tilde{\theta}_D$
	\end{tabular*}
\caption{Axial and vector couplings for SM quarks in a 2HDM scenario.}
\label{tab:VAChargesQ}
\end{center}
\end{table}

\begin{table}[t!]
\begin{center}
\begin{tabular*}{0.8\textwidth}{@{\extracolsep{\fill}}c|c|c|c}
\bf{SM neutrino} & $\nu_e$ & $\nu_\mu$ & $\nu_\tau$ \\
\hline
 $c^\nu \equiv c_V^\nu = c_A^\nu$ & $\frac{1}{2} g_D  \tilde{\theta}_D$ & $\frac{1}{2} g_D ( \tilde{\theta}_D +  \kappa)$ & $\frac{1}{2} g_D ( \tilde{\theta}_D - \kappa)$	\end{tabular*}
\caption{Neutrino couplings in a 2HDM scenario.}
\label{tab:VAChargesNu}
\end{center}
\end{table}

We further allow for kinetic mixing between $U(1)_D$ and $U(1)_Y$, parameterized by $\epsilon$. As discussed in section~\ref{sec:1HDM-kinmix}, such a module does not influence the axial couplings to leading order 
in the mixing parameter $\epsilon$; all instances of ${\cal O}(\epsilon)$  couplings arise from $\apr$ mixing with the SM photon and only affect the vector couplings between  $\apr$ and charged fermions. 

All together, we consider models parameterized by six quantities,
\be
\{g_D, q_{H_{u}}, q_{H_{d}}, \tilde{\theta}_D, \epsilon, \kappa \},
\ee
where the overall scale of the couplings is set by $g_D$;
the charges $q_{H_u}$, $q_{H_d}$ and mass mixing $\tilde{\theta}_D$ control the axial couplings; $\epsilon$ controls the relative size of the vector couplings; and $\kappa$ controls the muon vector coupling with respect to the electron. The SM fermion axial and vector couplings in the mass basis are summarized in tables \ref{tab:VAChargesL}, \ref{tab:VAChargesQ}, and \ref{tab:VAChargesNu}. 

\subsection{Dark Higgs Bosons}

We have remained somewhat agnostic about the properties of the additional Higgs bosons which are generically present when anomalons have $U(1)_D$ charges.  In a model with two SM Higgs doublets $H_u$ and $H_d$,
there is rich phenomenology that is relatively well-understood from studies of the minimal supersymmetric standard model.  Constraints from both
collider searches and precision measurements generically require that
these additional Higgs bosons have masses greater than several hundred GeV.  Dark Higgses $H'$ which are SM singlets are much less constrained since their interaction with the SM generically goes through the $A'$, and this coupling can be weak for small gauge coupling. For $m_{H'} > 2 m_{A'}$, the dominant decay mode of the $H'$ is $H' \to A' A'$, and for $m_{A'} < 2m_\mu$, searches for $e^+ e^- \to H' A'$ will have 3 $e^+ e^-$ pairs in the final state, which must compete with a large QED background \cite{Batell:2009yf}. Belle \cite{TheBelle:2015mwa} was able to set limits in the context of a kinetic mixing model corresponding to $\epsilon \lesssim 8 \times 10^{-4}$ for dark fine-structure constant $\alpha_D = 1/137$ and $m_{H'} < 8 \ \GeV$, but only for $m_{A'} > 100 \ \MeV$.

The presence of axial couplings to fermions implies an upper limit on the mass of any Higgs contributing to 
the $A'$ mass -- including the SM Higgs $h^0$ as well as any dark Higgses.
Perturbative unitarity dictates that the masses of these Higgs bosons satisfy \cite{Kahlhoefer:2015bea}:
\be
\label{eq:AxialUnitarityBound}
m_{H,H'} \lesssim \frac{\pi m_{A'}^2}{(c_A^f)^2 \, m_f},
\ee
which is driven by the mass of the heaviest fermion $f$ in the theory with axial charges under $\UD$. By the arguments above, a nonzero axial coupling to the up quark implies that Eq.~(\ref{eq:AxialUnitarityBound}) applies for the top quark. If the $A'$ gets its mass entirely from SM Higgses in \emph{e.g.} the 2HDM considered above, 
the bound is parametrically $m_H \lesssim v$, which is trivially satisfied for the SM Higgs $h^0$ but in tension with the heavier Higgses which have not been observed below the scale $v$. However, since the anomalons are generally constrained to be heavier than 1 TeV and get their mass from additional dark Higgses, these may have tighter mass bounds than the SM Higgs, providing an appealing target for GeV-scale dark sector searches. Indeed, the additional dark Higgs contributions to the $A'$ mass relax the tension with the heavy SM Higgses in Eq.~(\ref{eq:AxialUnitarityBound}).


\section{Surveying The Parameter Space}
\label{sec:Survey}

\begin{figure}[t!] 
\hspace{-0.9 cm} 
\includegraphics[width=0.528\textwidth]{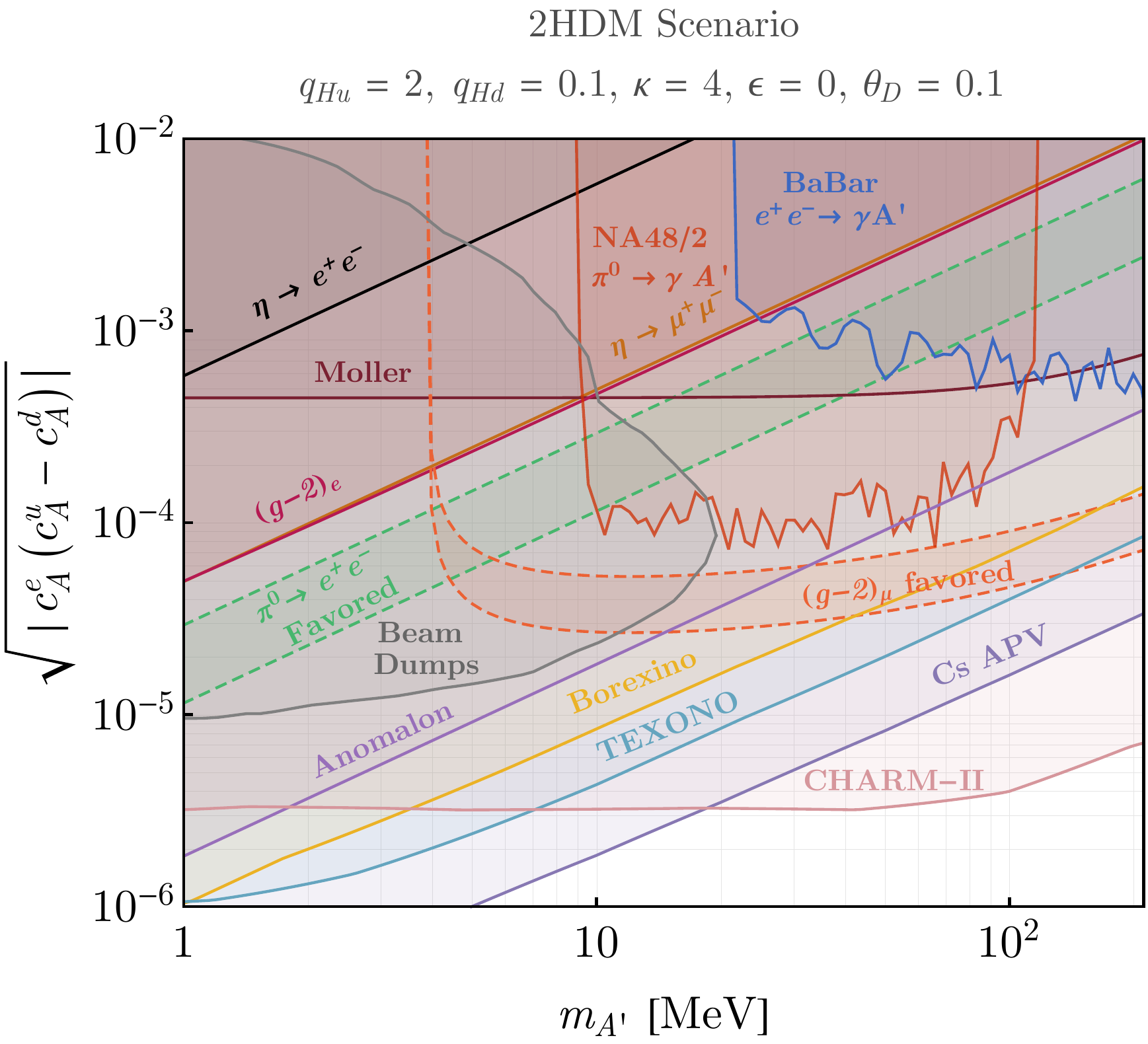} \hspace{-0.29 cm}~ 
\includegraphics[width=0.5185\textwidth]{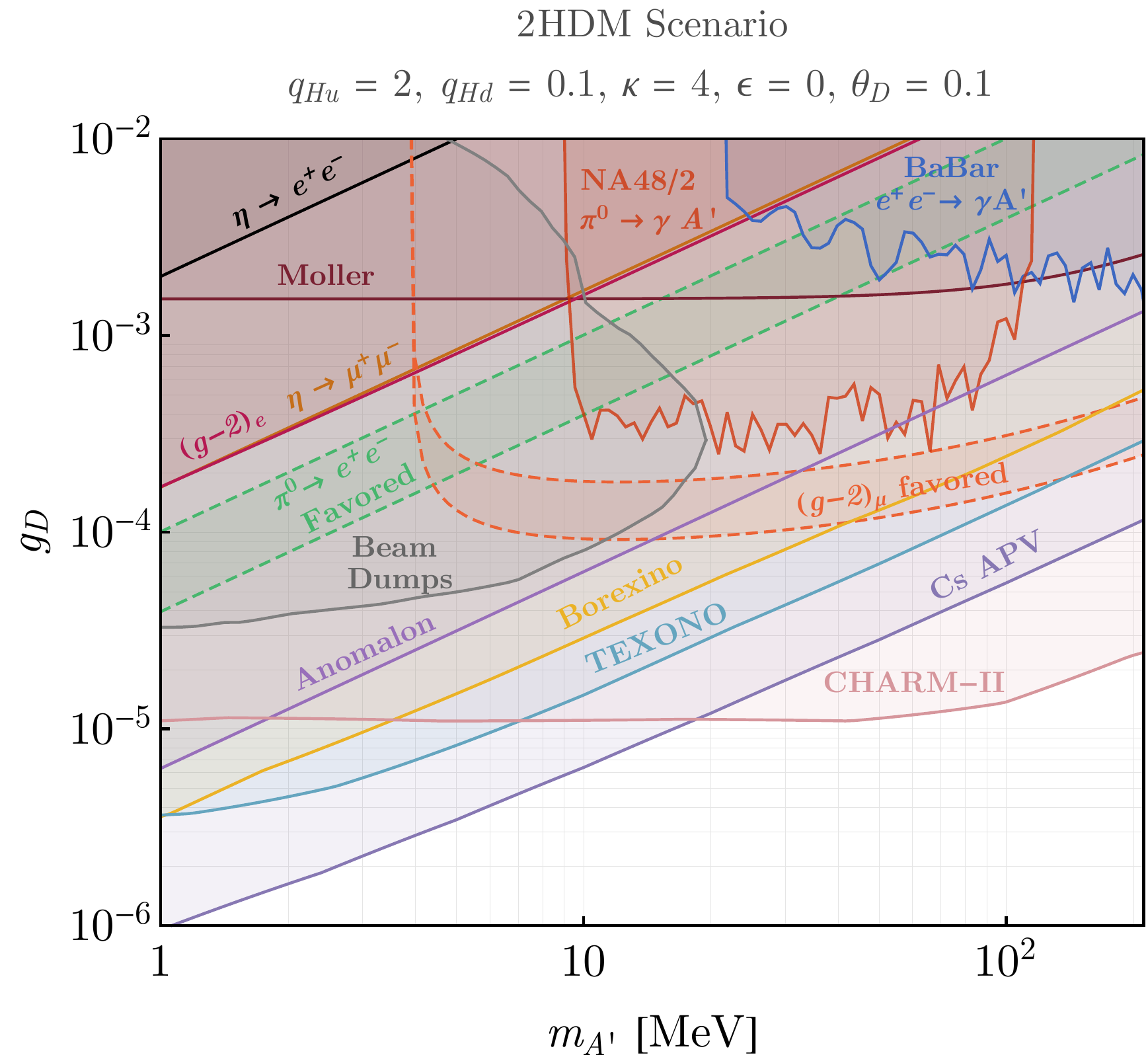} \hspace{-0.8cm}
\caption{Representative constraints on the parameter space for the 2HDM realization of axial couplings. The left panel plots the effective coupling for pseudoscalar decay $\sqrt{c_A^e (c_A^u - c_A^d)}$, while the right panel plots the gauge coupling $g_D$.} 
\label{fig:CouplingsGen} 
\end{figure}

Figure \ref{fig:CouplingsGen} shows a representative point in the parameter space of the Type-II 2HDM. A nonzero value of $\kappa$ was chosen for illustration in order to preserve the $(g-2)_\mu$ favored region; with $\kappa = 0$, the sizable axial coupling tends to push the anomalous magnetic moment below the SM value, turning the measured positive deviation into a constraint which excludes the majority of the displayed parameter space (see the discussion in section~\ref{sec:IRConstraints}). The left panel plots the effective coupling $\sqrt{c_A^e (c_A^u - c_A^d)}$ which controls the $A'$ contribution to the rare pseudoscalar decays $\pi^0 \to e^+ e^-$, $\eta \to e^+ e^-, \mu^+ \mu^-$ (see appendix \ref{app:Pseudoscalars}), while the right panel plots the overall coupling strength $g_D$. In principle, there are two distinct regions for $\pi^0 \to e^+ e^-$, arising from the fact that the SM contribution is 1-loop while the $A'$ contribution is tree-level; see appendix \ref{app:Pseudoscalars} for more details. For these particular values, there is only one region, which, interestingly, is consistent with $\eta \to \mu^+ \mu^-$ to $2\sigma$ for all $m_{A'}$.\footnote{This result corrects the conclusion of \cite{Kahn:2007ru}, where a simplified estimate suggested that the measured $\pi^0$ branching ratio was inconsistent with measurements of $\eta \to \mu^+ \mu^-$ for the same $A'$ mass and axial couplings.} Nonetheless, this region is in conflict with numerous constraints. At low masses, the most stringent constraint is a measurement of the weak charge of cesium in atomic parity-violation experiments \cite{Ginges:2003qt,Wood:1997zq}, and at higher masses, the $A'$ searches from BaBar \cite{Lees:2014xha} and NA48/2 \cite{Batley:2015lha} come into play. Constraints from neutrino scattering are also quite severe, and are dominated by the CHARM-II~\cite{Vilain:1994qy} $\overline{\nu}_\mu-e$ bounds above $m_{A'} \simeq 20 \ \MeV$, since the large value of $\kappa$ enhances the $\nu_\mu$ coupling with respect to $\nu_e$.\footnote{The neutrino-electron bounds here are translated from \cite{Jeong:2015bbi} by assuming $c_A^e \sim c_V^e$, $c^{\nu_\mu} \sim \kappa c^{\nu_e}$, which is true for a generic point in the 2HDM parameter space and suffices to illustrate the dominance of neutrino bounds for large $\kappa$.} The range of $m_{A'}$ and $g_D$ (or $c_A$) required to explain both $(g-2)_\mu$ and the pseudoscalar decays is robustly excluded by several independent measurements, at least for this point in parameter space.

The lines labeled ``anomalons'' represent the bound in Eq. (\ref{eq:anomalonconstraint}).  As discussed above, LHC searches for
heavy colored fermions bound their masses to be $\gtrsim 1 \ \TeV$. Since anomalon masses are generated by a
dark Higgs VEV which also contributes to $m_{A'}$, this bound can restricts the parameter space in the $m_{A'}-g_D$ plane even more severely than constraints from $(g-2)_e$. Indeed, for the parameters chosen in figure~\ref{fig:CouplingsGen}, the anomalon constraint excludes completely the preferred region for $\pi^0 \to e^+ e^-$, independent of any IR constraints. However, there is an interesting corner of parameter space consistent with the anomalon bound where $g_D \ll 1$ and the $A'$ is very light: for example, $y_\psi = 1$, $g_D = 10^{-12}$, $m_A' \sim 10 \ \eV$. Light dark matter charged under $\UD$ can scatter with electrons in a superconductor through the $A'$. Unlike in the case of purely vector coupling, an axially-coupled $A'$ will not acquire a keV-scale effective mass \cite{ZurekPrivate}, potentially enhancing the direct detection rate by several orders of magnitude compared to the estimates of \cite{Hochberg:2015fth}.

\begin{figure}[t!] 
\hspace{-1.1cm}
\includegraphics[width=0.52\textwidth]{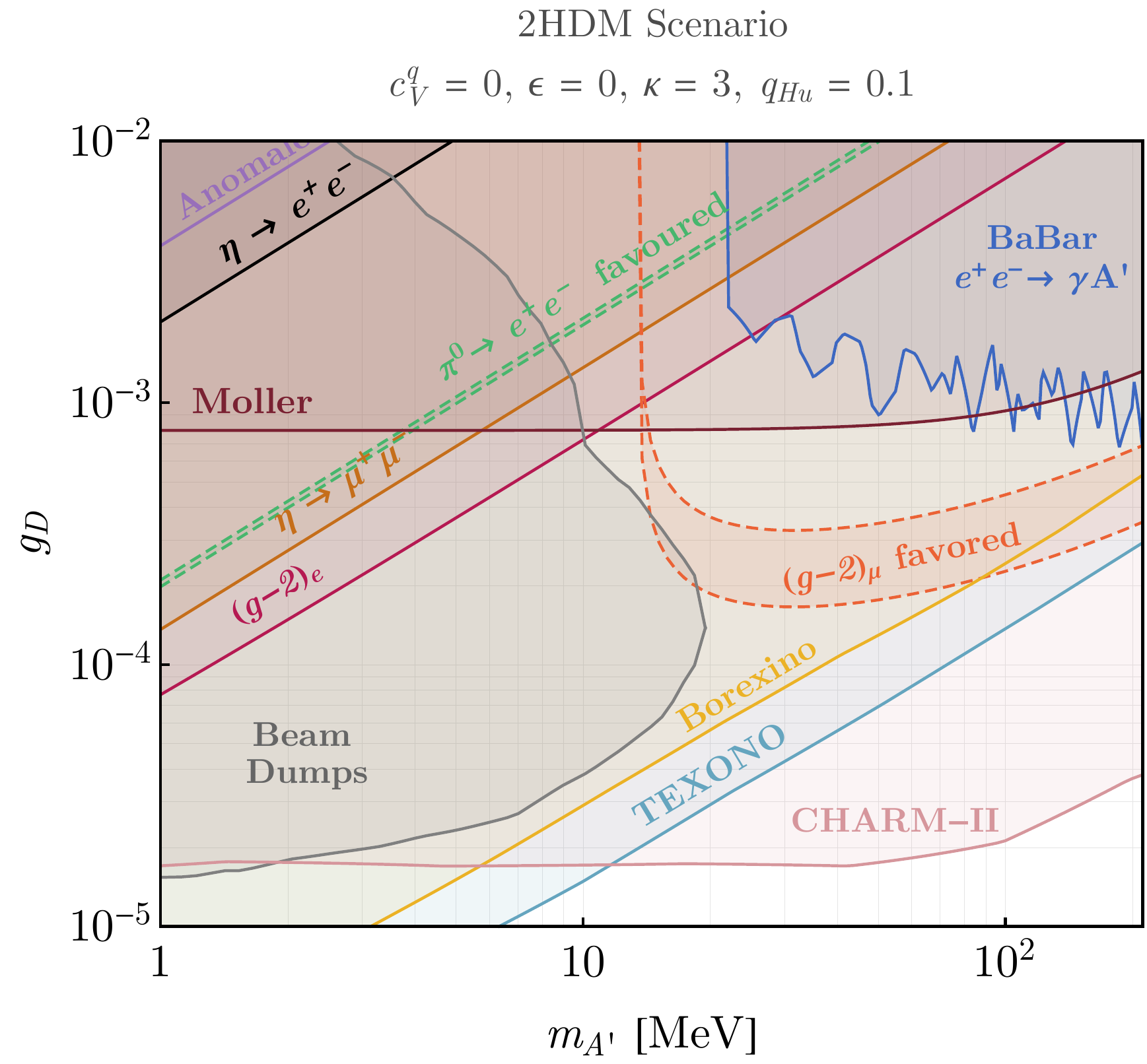} \hspace{-0.1 cm}~~~
\includegraphics[width=0.52\textwidth]{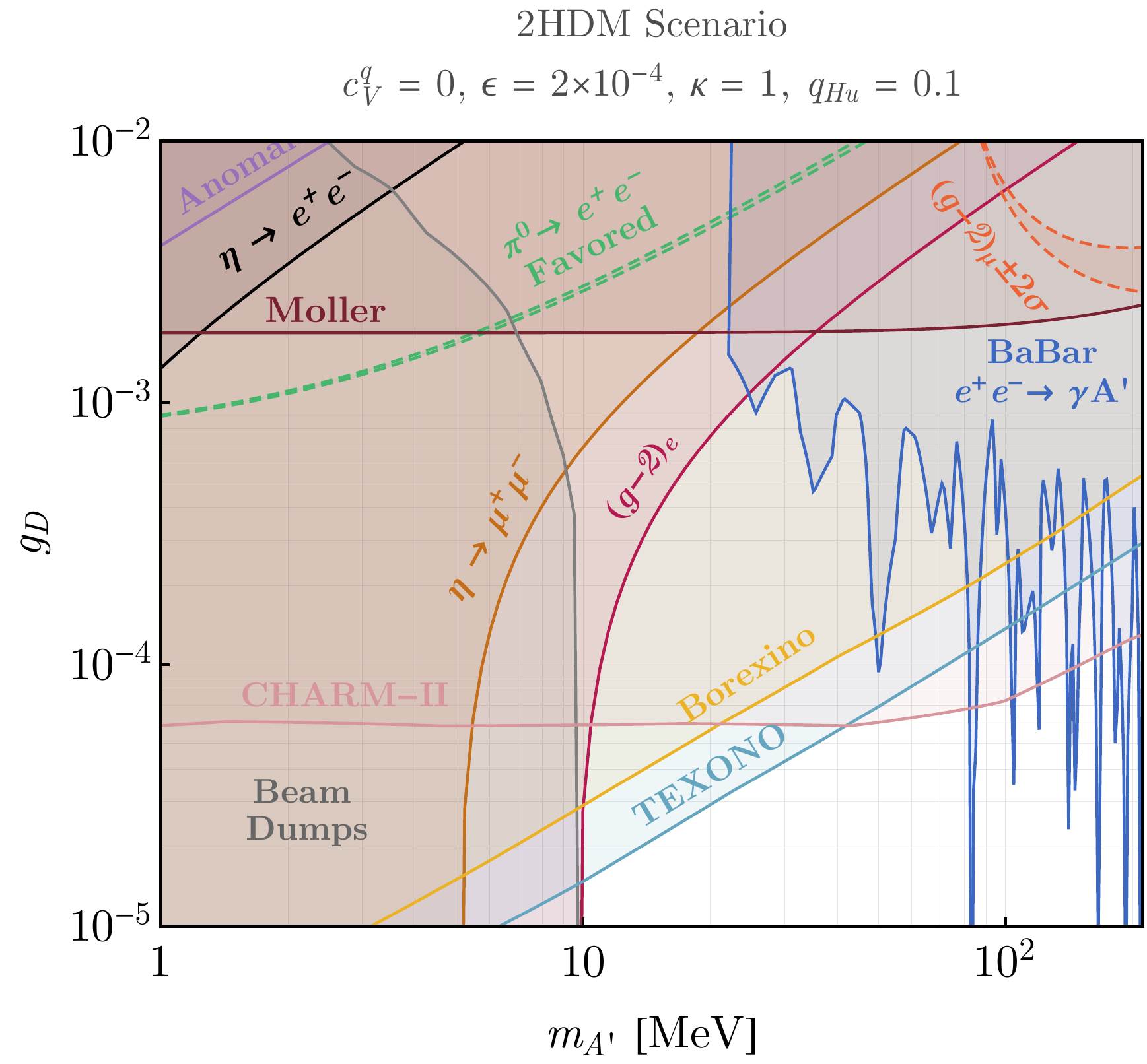} \hspace{-0.8cm}
\caption{Constraints on couplings in the 2HDM, with free parameters chosen so that $c_V^q = 0$ for all quarks, for different values of $\kappa$ and $\epsilon$.} 
\label{fig:CouplingsCv0} 
\end{figure}

As mentioned in the Introduction, one can attempt to evade the most stringent constraints by constructing a model where $c_V^q = 0$. However, simply choosing $c_V = 0$ is in tension with gauge invariance, as the models of section \ref{sec:GaugeInvtYukawas} have shown. In the Type-II 2HDM, it is possible to fine-tune both $c_V^u$ and $c_V^d$ to zero, but only at the expense of reintroducing other couplings. Indeed, table \ref{tab:VAChargesQ} shows that setting $c_V^u = c_V^d = 0$ will fix $\epsilon$ and $\tilde{\theta}_D$ in terms of $q_{H_u}$, $q_{H_d}$, and $g_D$, and a nonzero value of $\tilde{\theta}_D$ necessarily implies nonzero neutrino couplings, as shown in table \ref{tab:VAChargesNu}. The parameter space for this fine-tuned scenario is shown in figure~\ref{fig:CouplingsCv0} for different choices of $\epsilon$ and $\kappa$. For $\epsilon = 0$ (Fig~\ref{fig:CouplingsCv0}, left), the neutrino constraints from TEXONO \cite{Davoudiasl:2014kua} are dominant, while for nonzero epsilon and small $\kappa$ (Fig~\ref{fig:CouplingsCv0}, right), $(g-2)_e$ and BaBar dominate in some regions but the $(g-2)_\mu$ preferred region is pushed even deeper into other exclusion regions. For both parameter points, although anomalon bounds are weaker than in figure~\ref{fig:CouplingsGen}, the entire preferred region for $\pi^0$ decay is now excluded by several independent measurements, including $\eta \to \mu^+ \mu^-$.

\section{Loopholes from Flavor Breaking}
\label{sec:NonRen}

We have found that theoretical considerations place interesting constraints on the couplings of a light axially coupled
gauge boson, such that the theory remains valid up to
TeV-scale energies.  One of the key assumptions which underlies these conclusions is our choice to generate
axial-vector interactions for the SM fermions while still allowing for renornalizable Standard Model Yukawa interactions.  In this section, we
step away from this assumption, and consider in turn theories in which the SM fermions are uncharged under $U(1)_D$, but pick up small coupling to its
gauge boson through $U(1)_D$-breaking effects; theories in which the SM Yukawa interactions are effectively higher-dimensional operators; and
theories in which $U(1)_D$ acts in a family-dependent manner.

\subsection{Mixing with Vector-Like Fermions}

It is possible to engineer small couplings of $A^\prime$ to the SM fermions without charging them under $U(1)_D$, but 
inducing an interaction by
mixing with a set of vector-like fermions, which are charged under the new force.  To avoid constraints from precision measurements, 
such fermions should be in the same $SU(3)_c \times SU(2)_L \times U(1)_Y$ representations as the SM fermions with
which they mix, and should have the same charge under $U(1)_D$ as a dark Higgs $H'$ which is a SM singlet.  
For example, coupling to the up-type quarks can be engineered by introducing
vector-like states ${\cal U}_i$ and ${\cal Q}_i$ transforming under $(SU(3)_c, SU(2)_L, U(1)_Y, U(1)_D)$ as
$({\bf 3}, {\bf 1}, +2/3, a)$ and $({\bf 3}, {\bf 2}, +1/6, a)$, respectively, along with fields with conjugate
representations \:${\cal U}_i^c$ and ${\cal Q}_i^c$.  The index $i=1...3$ is a flavor index:
following the principle of minimal flavor violation (MFV) \cite{DAmbrosio:2002ex}, we construct each vector-like fermion as a triplet under its corresponding
$SU(3)$ subgroup of the $SU(3)^5$ flavor symmetry of the SM in the limit of vanishing Yukawa interactions.  MFV dictates that all breaking of $SU(3)^5$ be proportional to
the Yukawa matrices themselves,  insuring modest contributions to the most constraining flavor-violating observables involving the first two generations.
In the example at hand, ${\cal Q}_i$ and ${\cal U}_i$ are each triplets under $SU(3)_Q$ and $SU(3)_u$, respectively.

The Lagrangian,
\begin{eqnarray}
{\cal L} = - M_{\cal Q} ~ {\cal Q}_i^c {\cal Q}_i - M_{\cal U} ~{\cal U}_i^c {\cal U}_i - y_L ~H' ~{\cal Q}_i^c Q_i - y_R ~H' ~{\cal U}_i u_i^c,
\end{eqnarray}
includes masses for the vector-like quarks and also induces mixing with the left- and right-handed SM up quarks.  The family-universal choice of mass parameters
$M_{\cal Q}$ and $M_{\cal U}$ and couplings $y_L$ and $y_R$ represent the leading MFV terms.
The mixing parameters for the left- and right-handed up quarks are given by,
\begin{eqnarray}
\theta_L \sim \frac{y_L~v'}{M_{\cal Q}}, ~~~~~~~ \theta_R \sim \frac{y_R~v'}{M_{\cal U}},
\end{eqnarray}
where $v' \equiv \langle H' \rangle$ is the VEV of the dark Higgs, which we assume is much less than either
$M_{\cal Q}$ or $M_{\cal U}$.  Through these mixings, the SM up-type quarks generically pick up both vector and axial-vector
couplings to $A^\prime$,
\begin{eqnarray}
c^u_V = -\frac{1}{2} a g_D \left(  \theta_L^2 + \theta_R^2  \right), ~~~~~
c^u_A = -\frac{1}{2} a g_D \left(  \theta_L^2 - \theta_R^2  \right),
\end{eqnarray}
which in the MFV limit are approximately family universal.
One can replicate this structure, introducing vector-like fermions ${\cal D}$, ${\cal L}$, and ${\cal E}$,
to arrange for couplings to the down-type quarks and leptons as well.  Note that in this scenario, the 
usual SM Yukawa couplings are allowed at tree level since SM fermions are singlets under $U(1)_D$. 

There is a limit to the size of the couplings which can be induced through this mechanism, given the lower bound ($\gtrsim 1$~TeV)
on the masses of the ${\cal Q}$ and ${\cal U}$ states from
direct searches, the perturbative limit ($\lesssim 4 \pi$)
on the size of the Yukawa interactions $y_L$ and $y_R$, and the connection between the VEV $v'$ and the mass of the $A^\prime$,
$m_{A^\prime} \gtrsim ag_D v'$.  Assembling these together, the natural size for $c_V$ and $c_A$ are:
\begin{eqnarray}
c_V, c_A & \sim & \frac{1}{2} \frac{y^2}{a g_D}  \left( \frac{m_{A'}}{M} \right)^2
\sim \frac{10^{-8}}{ag_D} 
\times \left( \frac{y}{4\pi} \right)^2 \times \left( \frac{m_{A'}}{10~{\rm MeV}} \right)^2 \times \left( \frac{1~{\rm TeV}}{M} \right)^2,
\end{eqnarray}
where $y$ and $M$ refer generically to the strength of the Yukawa interactions and the masses of the vector-like fermions appropriate for the SM fermion
in question.  Note that for fixed $m_{A'}$, increased coupling $g_D$ and/or charge $a$
generically leads to a {\em smaller} effective $c_V$ and $c_A$ induced through this mechanism. These small axial couplings are generally too feeble to be consistent with the $\pi^0$ anomaly.

\subsection{Non-Renormalizable SM Yukawa Couplings}

An alternative construction unchains the SM Higgs charge from those of the left- and right-handed SM fermions, 
by realizing the SM Yukawa interactions as non-renormalizable higher-dimensional operators.  In its most extreme limit,
this allows one to induce an axial-vector coupling of $A'$ to the SM fermions without any associated mass mixing coming
from the charges of the electroweak Higgs doublet(s).

For example, consider a module consisting of the SM quark doublet
$Q$ with $U(1)_D$ charge $q_{Q_{L}}$, right-handed up quark $u^c$ with charge $q_{u_{R}}$, and SM Higgs doublet with charge
zero.  For general $q_{Q_{L}}$ and $q_{u_{R}}$, the Yukawa interaction $y_u H Q u^c$ is forbidden by $U(1)_D$.  It can be engineered
by introducing vector-like quark doublets ${\cal Q}_i$ with charge $-q_{u_{R}}$ (along with conjugate states ${\cal Q}_i^c$), and a dark Higgs
$H'_u$ of charge $a \equiv -q_{Q_{L}} - q_{u_{R}}$ which gets a VEV $v'$.  These assignments allow one to construct mixing through Lagrangian,
\begin{eqnarray}
{\cal L} = - M_{\cal Q} ~ {\cal Q}_i^c {\cal Q}_i - y ~H' {\cal Q}_i^c Q_i~ - ~H {\cal Q} ~ y' ~u^c,
\end{eqnarray}
where following MFV, ${\cal Q}_i$ is constructed as a flavor triplet of $SU(3)_Q$, $M_{\cal Q}$ and $y$ are family-universal, and $y'$ is
proportional to the up-type quark Yukawa matrix.
Integrating out the vector-like quarks results in an effective Yukawa interaction $H Q u^c$ with magnitude given by,
\begin{eqnarray}
y_{\rm eff} = y~ y'~\frac{v'}{M_{\cal Q}}.
\end{eqnarray}
This module is easily extended to provide masses for
the down-type quarks and leptons by including additional vector-like fermions.

There are strong constraints on the size of the axial couplings based on the need to realize the large ($\sim 1$) top Yukawa interaction together
with a light mass for the axial mediator.
These translate into
\begin{eqnarray}
\label{eq:ytconstraint}
| c^t_A | & \leq & \frac{1}{2} yy'~ \frac{m_{A'}}{M_{\cal Q}} 
\sim 10^{-3} \times \left( \frac{y~y'}{(4\pi)^2} \right) \times \left( \frac{m_{A'}}{10~{\rm MeV}} \right)
\times \left( \frac{1~{\rm TeV}}{M_{\cal Q}} \right) .
\end{eqnarray}
For perturbative Yukawa couplings ($y, y' \sim 1$), this restricts $|c_A| \lesssim 10^{-5}$ for $m_{A'} \sim 10$~MeV.  While strictly speaking this bound only applies
to the top quark, MFV constructions effectively impose it on all three generations. Again, the axial couplings generated in this model are too small to explain the $\pi^0$ anomaly unless the Yukawa couplings are at the boundary of perturbativity, $y, y' \sim 4\pi$.

\subsection{Family Non-Universal Couplings}

If one is willing to allow for fine-tuning in the masses and couplings of the vector-like quarks such that flavor-violating effects cancel out in the mass
basis, it is possible to relax these bounds.  For example, one could charge only the first generation right-handed fermions
under $\UD$, which avoids the stringent neutrino bounds discussed in section~\ref{sec:Survey}. Examining the SM Yukawa sector with flavor indices restored, 
\be
{\cal L}_{y \cal, {\rm SM}} =  (y_u)_{ij} H Q_i u^c_j  +   (y_d)_{ij} H^\dagger Q_i d^c_j   + (y_e)_{ij} H^\dagger L_i e^c_j  + {\rm h.c.},
\ee
we see that an insertion of a dark Higgs VEV is required for the $i1$ entries in each of the SM Yukawa matrices, but the larger entries of
the $2 \times 2$ lower right corner (as well as the $12$ and $13$ entries) are trivially $\UD$-invariant.  Thus, the restrictions on $c_A$ are no longer inherited from the top quark, leading
to a relaxation of the constraint of Eq.~(\ref{eq:ytconstraint}) by several orders of magnitude.

\begin{figure}[t!] 
\begin{center}
\includegraphics[width=0.6\textwidth]{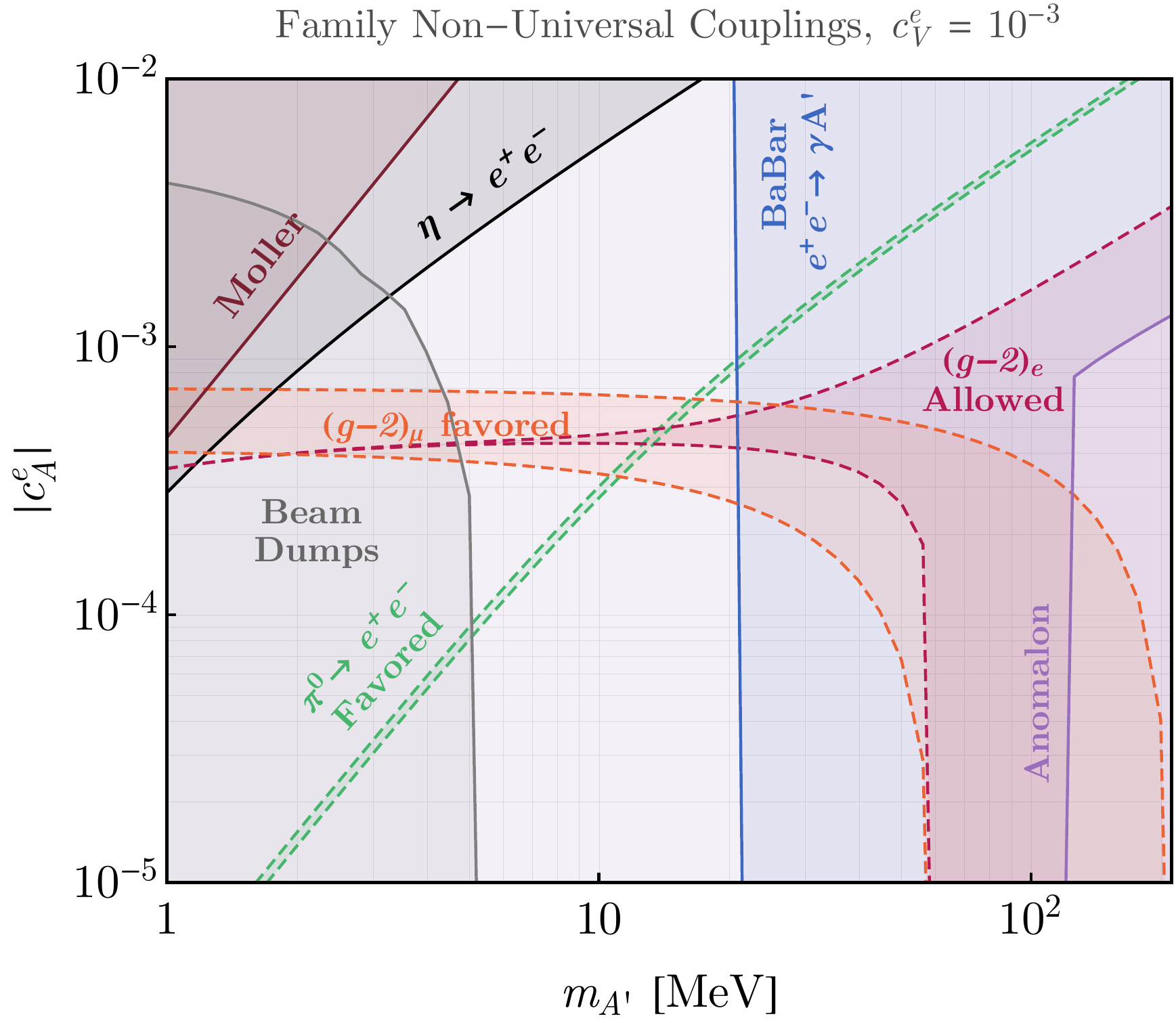}
\caption{Constraints on couplings with non-renormalizable Yukawa couplings, in the family non-universal case.} 
\label{fig:CouplingsFamilyTuned} 
\end{center}
\end{figure}

Provided one is willing to accept this (admittedly far-fetched) tuning, the theoretical constraints on this model are dominated by the generic anomalon bound, Eq.\ (\ref{eq:anomalonconstraint}). In this model, there are three dark Higgses $H'_u, H'_d, H'_e$, one for each Yukawa coupling we have to generate, with charges equal and opposite to the corresponding right-handed fermion charges. All three of these Higgses will contribute to the anomalon bound:
\be
m_{A'} \gtrsim g_D \sqrt{q_u^2 + q_d^2 + q_e^2} \times \left(\frac{4\pi}{y_\psi}\right).
\ee
Including kinetic mixing $\epsilon$, the first-generation SM fermion couplings are
\be
c_V^e  &=& -\frac{1}{2}g_D q_e - \epsilon e, \qquad \!~~~c_A^e = \frac{1}{2}g_D q_e, \\
c_V^u  &=& -\frac{1}{2}g_D q_u  + \frac{2}{3}\epsilon e, \qquad  c_A^u = \frac{1}{2}g_D q_u, \\
c_V^d   &=& -\frac{1}{2}g_D q_d - \frac{1}{3}\epsilon e, \qquad c_A^d = \frac{1}{2}g_D q_d.
\ee
Second- and third-generation fermions $f$ with electric charge $Q_f$ have vector couplings $c_V^f = \epsilon Q_f$ and vanishing axial couplings. As discussed in section\ \ref{sec:Survey}, the strongest experimental constraints can be evaded by setting $c_V^q = 0$ for first-generation quarks, which in this model effectively fixes $q_u$ and $q_d$ in terms of $\epsilon$, another fine-tuning. Neutrino couplings do not get generated from mass mixing with the $Z$ because the SM Higgs is uncharged. In figure \ref{fig:CouplingsFamilyTuned}, we plot the allowed parameter space for $c_A^e$ in this model for $c_V^e = 10^{-3}$. This model comes closest to realizing the generic IR parameter space described in section\ \ref{sec:IRParameterSpace} below $m_{A'} = 20 \ \MeV$ where BaBar loses sensitivity, albeit at the cost of several fine-tunings. Nonetheless, we see that the region compatible with both the $\pi^0 \to e^+ e^-$ and $(g-2)_\mu$ anomalies (which is also consistent with $(g-2)_e$) is now strongly excluded by the anomalon bounds, highlighting the tension between UV and IR considerations. Indeed, for this choice of $c_V^e$, the \emph{entire} parameter space in $c_A^e$ is ruled out by a combination of IR limits (BaBar) and UV limits (anomalons).

\section{Conclusions}
\label{sec:Conclusion}

Dark force carriers at the MeV scale are a fascinating possibility for physics beyond the Standard Model. They allow for a richer dark matter sector, which includes relevant interactions that offer new opportunities in model-building and for which there may even be experimental hints.  A large body of work has
focused on the case of vector interactions with the SM fermions, but it is worthwhile to understand the space of axially-coupled particles as well.
The chiral nature of the SM implies that realizing large axial couplings is non-trivial, with the shape of the IR physics impacted by UV physics living
at the TeV scale or above.

We have examined light force carriers with axial-vector interactions from both ends of the energy spectrum: from the low energy experimental perspective, where a rich set of constraints
from many searches provide complementary information, and also from the point of view of TeV models, to understand how the need for gauge invariance
under the full $SU(3)_c \times SU(2)_L \times U(1)_Y \times \UD$ impacts the phenomena that can be realized at MeV scales.  An immediate question is how
to reconcile the SM Yukawa interactions with the $\UD$ symmetry.  Models for which the charge assignments allow the SM Yukawas to be
realized at the renormalizable level are subject to interesting and subtle constraints.  For example, a model with a single Higgs doublet turns out to be
unable to realize large axial couplings below the electroweak scale because of a cancellation between the couplings inherited from $\UD$ and those induced by $A'$ mass mixing with
the $Z$.  In contrast, models with multiple Higgs bosons can evade this cancellation and can realize large axial-vector interactions, but mass mixing and the associated neutrino couplings are generic.  Orthogonal directions in theory space, in which the SM Yukawas are realized by integrating out messenger fermions, remain subject to restrictions on their parameters.
And in all cases, the need for additional matter to cancel anomalies allows searches for new particles at the LHC to shape the available parameters at
the MeV scale.

The coming years will see a host of new experiments seeking to map out the territory of dark vector particles.  Moving forward, it is important to remember that axial interactions are an interesting dimension of that space to explore. In particular, experiments which are only sensitive to the axial-vector coupling to SM leptons, such as improved measurements of the $\pi^0 \to e^+ e^-$ or $\eta \to \mu^+ \mu^-$ branching ratios, would be highly complementary to bremsstrahlung or beam dump experiments which measure a combination of vector and axial-vector couplings, as well as parity-violating observables like M{\o}ller scattering which measure products of vector and axial-vector couplings. At the same time, new observations, such as \emph{e.g.} the recently observed internal pair conversion of excited $^8$Be, may benefit from new force carriers with parity-violating interactions, and thus understanding the constraints on such theories can provide crucial information on the viable parameter space.

 \section*{Acknowledgments}
 We thank Daniele Alves, Paddy Fox, Susan Gardner, Ahmed Ismail, Andrey Katz, Zhen Liu, David Pinner, 
 Maxim Pospelov, Flip Tanedo, Jesse Thaler, and Kathryn Zurek
 for helpful conversations. Fermilab is operated by Fermi Research Alliance, LLC, under Contract No. DE-AC02-07CH11359 with the US Department of Energy.
TMPT is partially supported by NSF Grants PHY-1316792 and 1620638.
This work was partly completed at the Aspen Center for Physics, which is supported by National Science Foundation grant PHY-1066293. 
TMPT and GK would like to express a special thanks to the Mainz Institute for Theoretical Physics (MITP) for its hospitality and support during part of the time
in which this work was completed.\\

\noindent \textbf{Note Added:} In the final stages of preparing this article, \cite{Ismail:2016tod} appeared which addresses some similar issues in the context of theories with TeV-scale $A'$ masses.

\appendix

\section{Taxonomy of Constraints}
\label{app:Taxonomy}

In this appendix we briefly describe the relevant experimental constraints and the combinations of axial and vector couplings they depend on. A description of pseudoscalar decays, which motivated this study, can be found in appendix~\ref{app:Pseudoscalars}.

\begin{itemize}
\item $(g-2)_{\mu}$: The muon anomalous magnetic moment $(g-2)_{\mu}$ has had a persistent $\sim3\sigma$ discrepancy between the measured value~\cite{Bennett:2006fi,Mohr:2008fa} and SM prediction~\cite{Davier:2010nc,Hagiwara:2011af,Blum:2013xva}. It has long been recognized that a dark photon with vector couplings can contribute to this (positive) discrepancy; axial-vector couplings contribute to $(g-2)_{\mu}$ with the opposite sign. The total contribution is \cite{Fayet:2007ua}
\be
\delta a_\mu = \frac{(c_V^\mu)^2}{4\pi^2} \int \frac{x^2(1-x)}{x^2 + \frac{m_{A'}^2}{m_\mu^2}(1-x)}\,dx - \frac{(c_A^\mu)^2}{4\pi^2} \frac{m_\mu^2}{m_{A'}^2} \int \frac{2x^3 + (x-x^2)(4-x)\frac{m_{A'}^2}{m_\mu^2}}{x^2 + \frac{m_{A'}^2}{m_\mu^2}(1-x)}\,dx.
\label{eq:deltaamu}
\ee
Setting this equal to the observed deviation from the SM prediction produces a combination of $c_V^\mu$ and $c_A^\mu$ favoured by measurement.
\item $(g-2)_e$ : Vector and axial-vector electron couplings of dark photons contribute to the anomalous magnetic moment of the electron $(g-2)_e$ analogously to Eq.~(\ref{eq:deltaamu}):
\be
\delta a_e = \frac{(c_V^e)^2}{4\pi^2} \int \frac{x^2(1-x)}{x^2 + \frac{m_{A'}^2}{m_e^2}(1-x)}\,dx - \frac{(c_A^e)^2}{4\pi^2} \frac{m_e^2}{m_{A'}^2} \int \frac{2x^3 + (x-x^2)(4-x)\frac{m_{A'}^2}{m_e^2}}{x^2 + \frac{m_{A'}^2}{m_e^2}(1-x)}\,dx.
\ee
We require a combination of couplings $c_V^e$ and $c_A^e$ such that this value is consistent with the measured value~\cite{Hanneke:2008tm} given the SM prediction~\cite{Aoyama:2012wj}.
\item $e^+ e^- \to \gamma A', A' \to \ell^+ \ell^-$: BaBar looked for the production of dark photons through electron-positron annihilation ($e^+ e^- \to \gamma A'$) followed by decay of the dark photon into a charged lepton pair ($A' \to \ell^+ \ell^-,~\ell=e,\mu$)~\cite{Lees:2014xha}. The ordinary kinetic mixing case can be reinterpreted to constrain a combination of vector and axial electron couplings by constructing an effective $\epsilon$,
\be
e\epsilon_{\text{eff}}=\sqrt{(c_V^e)^2 + (c_A^e)^2}.
\label{eq:epseff}
\ee
\item  $\pi^0 \to \gamma A'$, $A' \to e^+ e^-$: The decay $\pi^0 \to \gamma A'$ proceeds through a mixed anomaly of the axial isospin current with $U(1)_{\rm EM} \times U(1)'$. Relative to the ordinary kinetic mixing case, we can compare with existing constraints from the NA48/2 experiment~\cite{Batley:2015lha} by constructing an effective $\epsilon$,
\be
\epsilon_{\rm eff} (Q_u^2 - Q_d^2) = Q_u c_V^u - Q_d c_V^d.
\ee
Note that this is also a dominant process for $A'$ production in proton beam dump experiments.

\item \emph{Atomic parity violation in Cesium}: $A'$ couplings to fermions induce a shift in the weak nuclear charge. Measurements of the weak nuclear charge of Cesium~\cite{Ginges:2003qt,Wood:1997zq} provide bounds on a combination of $c_A^e, c_V^u$ and $c_V^d$ when compared with the SM theoretical expectation. The $A'$ contribution to the weak charge is
\be
\Delta Q_{W} = -\frac{2\sqrt{2}}{G_\text{F}}c_{A}^{e}\left[\frac{c_{V}^{u}(2Z+N)+c_{V}^{d}(Z+2N)}{m_{A'}^{2}}\right]K(m_{A'}),
\ee
where $K(m_{A'})$ is an atomic form factor which accounts for the Yukawa-like potential involved in the $A'$-mediated interaction between the nucleus and electrons \cite{Bouchiat:2004sp}.
\item \emph{Parity-violating M{\o}ller scattering}: $A'$-electron axial and vector couplings contribute to the left-right asymmetry of electron-electron (M{\o}ller) scattering, $A_{PV} \equiv \frac{\sigma_L - \sigma_R}{\sigma_L+\sigma_R}$.  

The leading order parity violating process comes  from interference between
QED and $A'$ diagrams. For the QED amplitude with incident momenta $p_{i}$ and outgoing momenta $k_i$, the  left polarized contributions from $t$ and $u$ channel diagrams are 
\be
{\cal A}^L_{SM}  =\frac{e^2}{t}  [ \bar u_{k_1} {\ga}^\mu P_L u_{p1}  ][ \bar u_{k_2} {\ga}_\mu  u_{p_2}  ] - \frac{e^2}{u}  [ \bar u_{k_2} {\ga}^\mu P_L u_{p1}  ][ \bar u_{k_1} {\ga}_\mu  u_{p_2}  ] \equiv {\cal A}^L_{SM, t} - {\cal A}^L_{SM, u},
\ee
and the relative minus is from Fermi statistics. Similarly the $\apr$ exchange diagrams give contributions from the axial terms where 
\begin{align}
{\cal A}^L_{D} &  =\frac{c_A^2}{t-m_{\apr}^2}  [ \bar u_{k_1} {\ga}^\mu \ga^5 P_L u_{p1}  ][ \bar u_{k_2} {\ga}_\mu \ga^5 u_{p_2}  ] - \frac{c_A^2}{u-m_{\apr}^2}  [ \bar u_{k_2} {\ga}^\mu \ga^5 P_L u_{p1}  ][ \bar u_{k_1} {\ga}_\mu  \ga^5 u_{p_2}  ] \\
& \equiv {\cal A}^L_{D, t} - {\cal A}^L_{D, u}.
\end{align}
The additional contribution then comes from evaluating 
\be
\frac{d \sigma_{L,R} }{dE_e}\simeq \frac{m_e }{32 \pi s |\vec p_{cm}|} \langle          {\cal A}_{D} {\cal A}_{SM}^* + {\cal A}^*_{D} {\cal A}_{SM}   \rangle_{L,R}
\ee
at  each polarization. The PV scattering constraint comes from the SLAC E158 measurement of $A_{PV}$~\cite{Anthony:2005pm} at $Q^2 = 0.026 {\rm ~GeV}^2$, corresponding to $E_e \simeq 26$ GeV in the lab frame. Comparing this to SM expectation constrains the combination $c_{V}^{e}c_{A}^{e}$.

\item \emph{Neutrino-electron scattering}: If present, $A'$-neutrino couplings give rise to contributions to $\nu$-$e^{-}$ scattering involving combinations of $c_{V}^{e}c^{\nu}$ and $c_{A}^{e}c^{\nu}$, which can be constrained by existing measurements~\cite{Bilmis:2015lja}. The Borexino~\cite{Bellini:2011rx} ($\nu_{e}$-$e^{-}$), TEXONO~\cite{Davoudiasl:2014kua} ($\bar\nu_{e}$-$e^{-}$) and CHARM-II~\cite{Vilain:1994qy} ($\nu_{\mu}$-$e^{-}$) experiments provide the strongest constraints.

\item \emph{Electrom beam-dump experiments}:  These look for $A'$ production through brehmsstrahlung from electrons scattering off target nuclei, followed by their decay to leptons. Experiments E774~\cite{Bross:1989mp} at Fermilab and E141~\cite{Riordan:1987aw} at SLAC provide constraints in our parameter range of interest for $m_{A'} \lesssim 10$ GeV. Limits on the kinetic mixing $\epsilon$~\cite{Andreas:2012mt} can be translated using eq.~(\ref{eq:epseff}) to constrain the combination $(c_V^e)^2 + (c_A^e)^2$.
\end{itemize}

\section{Pseudoscalar Decays Revisited}
\label{app:Pseudoscalars}

In this appendix, we revisit the calculation of \cite{Kahn:2007ru} for the $A'$ contribution to $\pi^0 \to e^+ e^-$ in somewhat more detail, and extend the analysis to the case of rare $\eta$ decays.

\subsection{$\pi^0 \to e^+ e^-$}

The measured branching ratio for this process is $\Br(\pi^0 \to e^+e^-)_{\rm meas} = 7.48(38) \times 10^{-8}$ \cite{Abouzaid:2006kk}.\footnote{We note that this measurement is now 10 years old and relies on extrapolating the radiative tail of the $e^+ e^- \gamma$ final state.} An earlier calculation \cite{Dorokhov:2007bd} and a more recent model-independent one \cite{Masjuan:2015lca} both give SM values which are lower by at least $2\sigma$: $\Br(\pi^0 \to e^+e^-)_{\rm SM} \simeq 6.20 - 6.35 \times 10^{-8}$. As was pointed out in \cite{Kahn:2007ru}, the $A'$ contribution can bring the branching ratio into agreement with the measured value for appropriate axial couplings to quarks and leptons. 
However, some care is required because the SM contribution is one loop higher than the $A'$ contribution, so interference effects are important.

The leading-order SM contribution to $\pi^0 \to e^+ e^-$ is through a loop with two virtual photons. The matrix element can be written
\be
i\mathcal{M}_{SM}(q^2) = -e^2 f_{\pi \gamma \gamma} \int \frac{d^4 k}{(2\pi)^4} \frac{\epsilon_{\mu \nu \sigma \tau}k^\sigma q^\tau \tilde{F}_\pi(k^2, (q-k)^2)}{k^2(q-k)^2}  L^{\mu \nu},
\ee
where $q$ is the pion momentum, $k$ and $q-k$ are the photon loop momenta,
\be
L^{\mu \nu} = ie^2 4\sqrt{2}\frac{m_e}{m_\pi}\epsilon^{\mu \nu \alpha \beta}k_\alpha q_\beta \frac{1}{(k-p)^2 - m_e^2}
\ee
is the spin-singlet projection of the lepton half of the diagram (lepton momenta $p$ and $q-p$), and
\be
f_{\pi \gamma \gamma} = \frac{1}{4\pi^2 f_\pi}
\ee
is the coupling to two real photons \cite{Bergstrom:1982zq,Masjuan:2015lca}. This normalizes the pion transition form factor $\tilde{F}_\pi$ to $\tilde{F}_\pi(0, 0) = 1$. Contracting the lepton tensor, we can parameterize the amplitude for on-shell pion decay as
\be
\label{eq:SMpiee}
i \mathcal{M}_{\rm SM}(m_\pi^2) = \frac{\sqrt{2}m_e m_\pi \alpha^2}{\pi^2 f_\pi} \mathcal{A}(m_\pi^2),
\ee
where
\be
\mathcal{A}(q^2) = 2i \int \frac{d^4 k}{\pi^2} \frac{(q^2 k^2 - (q \cdot k)^2)\tilde{F}_\pi(k^2, (q-k)^2)}{q^2 k^2 (q-k)^2 ((p-k)^2 - m_e^2)}
\ee
is the loop integral calculated in \cite{Masjuan:2015lca}.

We can write the tree-level contribution through a virtual $A'$ as
\be
i\mathcal{M}_{A'}(q^2) =\left(\frac{-i(g_{\mu \nu} - q_\mu q_\nu/m_{A'}^2)}{q^2 - m_{A'}^2}\right)  (i c_A^e L^\mu) i  \langle 0 | \overline{Q}  c_A^Q \gamma^\nu \gamma^5 Q | \pi(q) \rangle,
\ee
where the lepton tensor is
\be
L^\mu = [\overline{u}(p)\gamma^\mu \gamma^5 v(p')]_{\rm spin-0} = 2\sqrt{2}\, \frac{m_e}{m_\pi} q^\mu
\ee
and $q = p + p'$.
Note that axial-vector couplings are required on both sides of the diagram: the spin-zero component of the purely vector lepton tensor $\overline{u}\gamma^\mu v$ vanishes. Anticipating the 3-flavor case we will need for the $\eta$, we use $Q = (u, d)^T$ and the matrix of axial couplings $c_A^Q = {\rm diag}(c_A^u, c_A^d)$. Now, any $2 \times 2$ matrix can be written as a linear combination of Pauli matrices and the identity, so
\be
c_A^Q = k_a \tau^a
\ee
where $\tau^0$ is the identity and $\tau^i$, $i = 1, 2, 3$ are the Pauli matrices. The orthogonality relation $\Tr(\tau^a \tau^b) = 2 \delta^{ab}$ implies in particular
\be
k_3 = \frac{1}{2} \Tr(c_A^Q \tau^3) = \frac{1}{2}(c_A^u - c_A^d).
\ee
From the definition of the axial isospin current, 
\be
\langle 0 |  \overline{Q}  \gamma^\nu \gamma^5 \frac{\tau^j}{2} Q | \pi^k (q) \rangle = i \delta^{jk} f_\pi q^\nu,
\ee
only the third component has overlap with the pion, and so finally 
\be
\langle 0 | \overline{Q}  c_A^Q \gamma^\nu \gamma^5 Q | \pi(q) \rangle = i (c_A^u - c_A^d) f_\pi q^\nu.
\ee
This gives
\be
i \mathcal{M}_{A'}(m_\pi^2) = 2\sqrt{2} f_\pi m_e m_\pi \frac{c_A^e(c_A^u - c_A^d)}{m_{A'}^2}.
\ee
Adding this to (\ref{eq:SMpiee}) and squaring, after plugging in ${\rm Re} \, \mathcal{A}(m_\pi^2) = 10.0-10.46$ and ${\rm Im} \, \mathcal{A}(m_\pi^2) = -17.52 $ from \cite{Masjuan:2015cjl} one can set the branching ratio equal to the measured value and solve for the combination $c_A^e(c_A^u - c_A^d)/m_{A'}^2$. Note that the $A'$ amplitude contributes to the real part of $\mathcal{A}(m_\pi^2)$, and that there are potentially two real solutions to this quadratic equation. Also note that the $A'$ could contribute either constructively or destructively; any deviation from the SM prediction can in principle be explained by an $A'$ with suitable mass and coupling.

\subsection{$\eta \to \mu^+ \mu^-, e^+ e^-$}

According to the calculations of \cite{Masjuan:2015cjl}, $\Br(\eta \to \mu^+ \mu^-)_{\rm SM} = (4.52-4.72) \times 10^{-6}$, but $\Br(\eta \to \mu^+ \mu^-)_{\rm meas} = 5.8(8) \times 10^{-6}$, so in this case SM theory and experiment are consistent to about $1\sigma$. The above framework carries over almost identically for the $\eta$, the only difference being the replacement of the pion form factors with the appropriate singlet and octet values. In the SM amplitude, $f_\pi$ should be replaced by $\frac{1}{4\pi^2 f_{\eta \gamma \gamma}}$, where $f_{\eta \gamma \gamma} = 2.74(5) \times 10^{-4} \ \MeV^{-1}$ is related to the $\eta \to \gamma \gamma$ branching ratio by $f_{\eta \gamma \gamma}^2 = \frac{64\pi}{(4\pi \alpha)^2 m_\eta^3}\Gamma(\eta \to \gamma \gamma)$ \cite{Masjuan:2015cjl}. The SM amplitude is
\be
i \mathcal{M}_{\rm SM}(m_\eta^2) = 4 \sqrt{2}m_\mu m_\eta f_{\eta \gamma \gamma} \alpha^2 \mathcal{A}(m_\eta^2),
\ee
while the $A'$ contribution is
\be
i \mathcal{M}_{A'}(m_\eta^2) = \sqrt{\frac{8}{3}} m_\mu m_\eta \frac{1}{m_{A'}^2} c_A^\mu \left ( \sqrt{2}(c_A^u + c_A^d + c_A^s)f^0_\eta + (c_A^u + c_A^d - 2c_A^s)f^8_\eta \right).
\ee
There is considerable theoretical uncertainty in the decay constants. They can be parameterized as $f^0_\eta = -F_0 \sin \theta_0$ and $f^8_\eta = F_8 \cos \theta_8$, with $F_0 \simeq 115 \ \MeV$, $F_8 \simeq 120 \ \MeV$, $\theta_0 \simeq 0$, $\theta_8 \simeq -19^\circ$  \cite{Bickert:2015cia}. The near-vanishing of $\theta_0$ means that we can approximate the contribution to the decay as pure octet,
\be
i \mathcal{M}_{A'} = \sqrt{\frac{8}{3}} m_\mu m_\eta \frac{1}{m_{A'}^2} c_A^\mu (c_A^u + c_A^d - 2c_A^s) \widetilde{F}
\ee
where $\widetilde{F} \simeq 113 \ \MeV$. As discussed in section \ref{sec:1HDM}, for a renormalizable model of $A'$ interactions, gauge invariance of the SM Yukawa terms requires $c_A^d = c_A^s$, so this decay depends on the same combination of parameters $c_A^e(c_A^u - c_A^d)/m_{A'}^2$ as $\pi^0 \to e^+ e^-$. As before, one can plug in ${\rm Re} \, \mathcal{A}(m_\eta^2) = -(0.99-1.52)$ and ${\rm Im} \, \mathcal{A}(m_\eta^2) = -5.47$ to solve for this combination. 

The calculation for $\eta \to e^+ e^-$ is identical, \emph{mutatis mutandis}. This decay has not been observed, and indeed the best limits on the branching ratio \cite{Agakishiev:2013fwl,Moskal:2014dsa} lie more than three orders of magnitude above the SM unitarity bound of $\sim 10^{-9}$, so $\eta \to e^+ e^-$ only provides a rather weak constraint.

\bibliography{AxialDarkPhotonBib}

\end{document}